\RequirePackage{fix-cm} % Fix LaTeX2e bugs.
\documentclass[a4paper, twoside, reqno, dvips, 12pt]{amsart}
\usepackage{fixltx2e}   % Fix LaTeX2e bugs.

\usepackage[latin1]{inputenc}
\usepackage[T1]{fontenc}

\usepackage{eucal}
\usepackage{dsfont}
\usepackage{xspace}
\usepackage{amsgen}
\usepackage{amsthm}
\usepackage{amssymb}
\usepackage{amsmath}
\usepackage{upgreek}
\usepackage{amsfonts}
\usepackage{MnSymbol}
\usepackage{mathrsfs}
\usepackage{stmaryrd}
\usepackage{mathtools}
\usepackage[nice]{nicefrac}

\usepackage{a4wide}

\usepackage[dvipsnames, table]{xcolor}

\definecolor{oneblue}{rgb}{0.0, 0.0, 0.85}
\definecolor{bluepigment}{rgb}{0.2, 0.2, 0.6}
\definecolor{darkgrey}{rgb}{0.273, 0.281, 0.30}
\definecolor{Lightgray}{rgb}{0.89, 0.89, 0.89}
\definecolor{Lightblue}{RGB}{214, 214, 214}
\definecolor{bckg}{RGB}{20.8, 20.8, 20.8} % Color of the boxes
\definecolor{charcoal}{rgb}{0.21, 0.27, 0.31}
\definecolor{darkelectricblue}{rgb}{0.33, 0.41, 0.47}

\usepackage[sort&compress, comma, square, numbers]{natbib}

\usepackage{psfrag}
\usepackage{graphicx}
\usepackage{subfigure}
\usepackage{morefloats}
\usepackage{indentfirst}

\usepackage{acronym}
\usepackage{microtype}
\usepackage[labelsep=period,%
            labelfont={bf,sf,color=bluepigment},%
            justification=raggedright]{caption}

\usepackage[perpage, symbol]{footmisc}

\usepackage[usenames, dvipsnames, pdf]{pstricks}
\usepackage{epsfig}
\usepackage{pst-grad} % For gradients
\usepackage{pst-plot} % For axes

\usepackage[colorlinks,
           urlcolor=oneblue,
           linkcolor=oneblue,
           citecolor=NavyBlue,
           bookmarksopen=false,
           pdfpagemode=UseNone,
           pagebackref]{hyperref}

\usepackage[explicit]{titlesec}

\titleformat{\section}
  {\color{NavyBlue}\Large\sffamily\bfseries}
  {}
  {0em}
  {\colorbox{bckg!5}{\parbox{\dimexpr\linewidth-2\fboxsep\relax}{\centering\thesection. #1}}}
  [\vspace*{0.33em}]

\titleformat{name=\section,numberless}
  {\color{NavyBlue}\Large\sffamily\bfseries}
  {}
  {0.0em}
  {\colorbox{bckg!10}{\parbox{\dimexpr\linewidth-2\fboxsep\relax}{\centering#1}}}
  [\vspace*{0.33em}]

\titleformat{\subsection}
  {\color{NavyBlue}\large\sffamily\bfseries}
  {}
  {0.0em}
  {\colorbox{bckg!5}{\parbox{\dimexpr\linewidth-2\fboxsep\relax}{\centering\thesubsection. #1}}}
  [\vspace*{0.33em}]

\titleformat{name=\subsection,numberless}
  {\color{NavyBlue}\Large\sffamily\bfseries}
  {}
  {0em}
  {\colorbox{bckg!10}{\parbox{\dimexpr\linewidth-2\fboxsep\relax}{\centering#1}}}
  [\vspace*{0.33em}]

\titleformat{\subsubsection}
  {\color{bluepigment}\sffamily\normalsize\bfseries}
  {\thesubsubsection}
  {0.5em}
  {#1}
  [\vspace*{0.33em}]

\titleformat{\paragraph}[runin]
  {\color{bluepigment}\sffamily\small\bfseries}
  {}
  {0em}
  {#1}

\titlespacing{\section}{1.0em}{1.5em plus 2pt minus 2pt}%
{1.0em plus 2pt minus 2pt}[0em]
\titlespacing{\subsection}{1.0em}{1.5em plus 2pt minus 2pt}%
{1.0em}[0em]
\titlespacing{\subsubsection}{1.0em}{1.5em plus 2pt minus 2pt}%
{1.0em plus 2pt minus 2pt}[0em]

\usepackage{titletoc}

\contentsmargin{0.5em}
\newlength{\tocsep} 
\titlecontents{section}[\tocsep]
  {\addvspace{10pt}\bfseries\sffamily}
  {\contentslabel[\thecontentslabel]{\tocsep}}
  {}
  {\ \titlerule*[0.75pc]{.}\ \thecontentspage}
  []
\titlecontents{subsection}[\tocsep]
  {\addvspace{8pt}\sffamily}
  {\contentslabel[\thecontentslabel]{\tocsep}}
  {}
  {\ \titlerule*[0.5pc]{.}\ \thecontentspage}
  []
\titlecontents*{subsubsection}[\tocsep]
  {\addvspace{2pt}\footnotesize\sffamily}
  {}
  {}
  {\ \titlerule*[0.35pc]{.}\ \thecontentspage}
  [\\*]

\makeatletter
\def\@setauthors{%
  \begingroup
  \def\thanks{\protect\thanks@warning}%
  \trivlist
  \centering\footnotesize \@topsep30\p@\relax
  \advance\@topsep by -\baselineskip
  \item\relax
  \author@andify\authors
  \def\\{\protect\linebreak}%
  \textsc{\normalsize\textcolor{darkelectricblue}{\authors}}%
  \ifx\@empty\contribs
  \else
    ,\penalty-3 \space \@setcontribs
    \@closetoccontribs
  \fi
  \endtrivlist
  \endgroup
}
\def\@settitle{\begin{center}%
  \baselineskip14\p@\relax
    \bfseries
    \textsc{\Large\textcolor{charcoal}{\@title}}
  \end{center}%
}
\makeatother

\usepackage{enumitem}
\setlist[description]{%
  topsep=30pt,               % space before start / after end of list
  itemsep=5pt,               % space between items
  font={\bfseries\sffamily\color{NavyBlue}}, % if colour is needed
}

\usepackage{fancyhdr}
\usepackage{lastpage}

\newcommand*\Title{\textcolor{bluepigment}{[Variational principles for modeling water waves}}
\newcommand*\Authors{\textcolor{bluepigment}{D.~Clamond \& D.~Dutykh}}
\newcommand*{\plogo}{\textcolor{gray}{{\texttt{arXiv.org} / \textsc{hal}}}} % Generic publisher logo

\numberwithin{equation}{section}

\newcommand{\vs}{\tilde{v}}

\newcommand{\udelta}{\delta}

\newcommand{\ud}{\mathrm{d}}

\newcommand{\ue}{\mathrm{e}}

\renewcommand{\beta}{\upbeta}
\newcommand{\phim}{\bar{\phi}}

\newcommand{\mus}{\tilde{\mu}}
\newcommand{\nus}{\tilde{\nu}}

\newcommand{\phib}{\check{\phi}}
\newcommand{\vx}{\boldsymbol{x}}

\newcommand{\vzero}{\boldsymbol{0}}
\newcommand{\phis}{\tilde{\varphi}}

\newcommand{\vum}{\bar{u}}
\newcommand{\um}{\bar{\/u}}
\newcommand{\us}{\tilde{u}}
\newcommand{\vmum}{\bar{\mu}}
\newcommand{\nub}{\check{\nu}}
\newcommand{\der}{\mathfrak{d}}
\newcommand{\vu}{\boldsymbol{u}}
\newcommand{\vmu}{\boldsymbol{\mu}}
\newcommand{\vub}{\check{\boldsymbol u}}
\newcommand{\vus}{\tilde{\boldsymbol u}}
\newcommand{\vmub}{\check{\boldsymbol \mu}}
\newcommand{\vmus}{\tilde{\boldsymbol \mu}}
\newcommand{\fatminus}{\ \rule[0.7mm]{2mm}{0.3mm}\ }

\newcommand{\cf}{\emph{cf.}~}
\newcommand{\ie}{\emph{i.e.}}
\newcommand{\eg}{\emph{e.g.}}

\newcommand{\etal}{\emph{et al.}~}

\newcommand{\nab}{\boldsymbol \nabla}
\newcommand{\scal}{\boldsymbol{\cdot}}
\newcommand{\grad}{\boldsymbol{\nabla}}

\newcommand{\half}{{\textstyle{1\over2}}}
\newcommand{\quat}{{\textstyle{1\over4}}}
\newcommand{\third}{{\textstyle{1\over3}}}
\newcommand{\sixth}{{\textstyle{1\over6}}}

\newcommand{\twothird}{{\textstyle{2\over3}}}

\usepackage{acronym}
\acrodef{sgn}[SGN]{Serre--Green--Naghdi}
\acrodef{bvp}[BVP]{Boundary Value Problem}
\acrodef{NSWE}{Nonlinear Shallow Water Equations}

\begin{document}

\title[\Title]{Practical use of variational principles for modeling water waves}

\author[D. Clamond]{Didier Clamond$^*$}
\address{Universit\'e de Nice -- Sophia Antipolis, Laboratoire J. A. Dieudonn\'e, 
Parc Valrose, 06108 Nice cedex 2, France}
\email{diderc@unice.fr}
\urladdr{http://math.unice.fr/~didierc/}
\thanks{$^*$ Corresponding author}

\author[D.~Dutykh]{Denys Dutykh}
\address{LAMA, UMR 5127 CNRS, Universit\'e Savoie Mont Blanc, Campus Scientifique, 
73376 Le Bourget-du-Lac Cedex, France}
\email{Denys.Dutykh@univ-savoie.fr}
\urladdr{http://www.denys-dutykh.com/}

\keywords{water waves; variational principle; Lagrangian; Hamiltonian; relaxation; approximations}

%%% ------------------------------------------------------------------------ %%%

\begin{titlepage}
\thispagestyle{empty} % Remove page numbering on this page
\noindent
{\Large Didier \textsc{Clamond}}\\
{\it\textcolor{gray}{Universit\'e de Nice -- Sophia Antipolis, LJAD, France}}
\\[0.02\textheight]
{\Large Denys \textsc{Dutykh}}\\
{\it\textcolor{gray}{Universit\'e Savoie Mont Blanc, CNRS, LAMA, France}}
\\[0.16\textheight]

\vspace*{0.75cm}

\colorbox{Lightblue}{
  \parbox[t]{1.0\textwidth}{
    \centering\huge\sc
    \vspace*{0.99cm}
    
    \textcolor{bluepigment}{Practical use of variational principles for modeling water waves}
    
    \vspace*{0.7cm}
  }
}

\vfill % Whitespace between the title block and the publisher

\raggedleft     % Right-align all text
{\large \plogo} % Publisher and logo
\end{titlepage}

%%% ------------------------------------------------------------------------ %%%

\newpage
\maketitle
\thispagestyle{empty}

%%% ------------------------------------------------------------------------ %%%

\begin{abstract}

This paper describes a method for deriving approximate equations for irrotational water waves. The method is based on a `relaxed' variational principle, \ie, on a Lagrangian involving as many variables as possible. This formulation is particularly suitable for the construction of approximate water wave models, since it allows more freedom while preserving a variational structure. The advantages of this relaxed formulation are illustrated with various examples in shallow and deep waters, as well as arbitrary depths. Using subordinate constraints (\eg, irrotationality or free surface impermeability) in various combinations, several model equations are derived, some being well-known, other being new. The models obtained are studied analytically and exact traveling wave solutions are constructed when possible.

\bigskip
\noindent \textbf{\keywordsname:} water waves; variational principle; Lagrangian; Hamiltonian; relaxation; approximations \\

\smallskip
\noindent \textbf{MSC:} \subjclass[2010]{76B15 (primary), 76M30, 76M22 (secondary)}
\smallskip \\
\noindent \textbf{PACS:} \subjclass[2010]{47.35.Bb (primary), 47.35.Fg, 47.85.Dh (secondary)}

\end{abstract}

%%% ------------------------------------------------------------------------ %%%

\newpage
\tableofcontents
\thispagestyle{empty}

%%% ------------------------------------------------------------------------ %%%

\newpage
\section{Introduction}

The water wave problem in fluid mechanics has been known since more than two hundreds years \cite{Craik2004}. The classical mathematical formulation of surface gravity waves involves five equations: the irrotationality of the fluid flow, the fluid incompressibility, the bottom and the surface impermeabilities, and the surface isobarity \cite{Mei1989}. This system of equations cannot generally be solved exactly and, historically, the water wave theory has been developed by constructing various approximations. In shallow water, we have the equations of \textsc{Korteweg} and \textsc{de Vries} \cite{KdV}, \textsc{Boussinesq} \cite{bouss}, \textsc{Benjamin} \etal \cite{bona}, \textsc{Serre} \cite{Serre1953}, \textsc{Green} and \textsc{Naghdi} \cite{Green1976}, \textsc{Camassa} and \textsc{Holm} \cite{Camassa1993}, \textsc{Degasperis} and \textsc{Procesi} \cite{Degasperis1999}, and many other model equations. On finite depth and deep water, there is the celebrated nonlinear Schr\"odinger equation \cite{Mei1989} and the equations of \textsc{Dysthe} \cite{Dysthe1979}, \textsc{Trulsen} \etal \cite{Trulsen2000}, \textsc{Kraenkel} \etal \cite{Kraenkel2005}, among others. These equations are most often derived via some perturbation techniques and are thus valid for waves of small amplitude. Moreover, these equations are generally valid for a very limited range of the ratio {\em wavelength\//\/water depth\/} and for narrow-banded spectra. However, for many applications it is necessary to use models uniformly valid for all depths and which are accurate for large amplitudes. It is well-known in theoretical physics that variational formulations are tools of choice to derive such approximations when small parameter expansions are inefficient.

There are mainly two variational formulations for irrotational surface waves that are commonly used, namely the Lagrangian of \textsc{Luke} \cite{Luke1967} and the Hamiltonian of \textsc{Petrov}--\textsc{Zakharov} \cite{Petrov1964, Zakharov1968}. Details on the variational formulations for surface waves can be found in review papers, \eg, \cite{Radder1999, Salmon1988, Zakharov1997}. The water wave problem is also known to have the multi-symplectic structure \cite{Bridges1996}. These variational principles have been exploited, in different variants, to build analytical and numerical approximations, \eg, \cite{Balk1996, Kim2001} just to mention a few references.

Luke's Lagrangian assumes that the flow is exactly irrotational, \ie, the Lagrangian involves a velocity potential but not explicitly the velocity components. If in addition the fluid incompressibility and the bottom impermeability are satisfied identically, the equations at the surface can be derived from Zakharov's Hamiltonian \cite{Zakharov1968}. Thus, both principles naturally assume that the flow is exactly irrotational, as it is the case of the water wave problem formulation, but Zakharov's Hamiltonian is more constrained than Luke's Lagrangian. Luke's and Zakharov's variational formulations require that part or all of the equations in the bulk of the fluid and at the bottom are satisfied identically, while the remaining relations must be approximated \cite{Craig1993}. It is because the irrotationality and incompressibility are mathematically easy to fulfill, that they are chosen to be satisfied identically. Beside simplicity, there are generally no reasons to fulfill irrotationality and/or incompressibility instead of the impermeability or the isobarity of the free surface, for example. In this paper, we illustrate the benefit of releasing the constraints of exact irrotationality and incompressibility, since approximations of these relations are sufficient in most practical cases.

Variational formulations involving as few dependent variables as possible are often regarded as simpler \cite{Yahalom2008}. It is understandably tempting to solve exactly (\ie, analytically) as many equations as possible in order to `improve' the solution accuracy. This is not always a good idea, however. Indeed, numerical analysis and scientific computing know many examples when efficient and most used algorithms do exactly the opposite. These so-called {\em relaxation methods\/} --- \eg, pseudo-compressibility for incompressible fluid flows \cite{Kameyama2005} --- have proven to be very efficient for stiff problems. When solving numerically a system of equations, the exact resolution of a few equations does not necessarily ensure that the overall error is reduced: what really matters is that the global error is minimized. A similar idea of relaxation may also apply to analytical approximations.

In this study we would like to elucidate the benefit of using relaxed variational methods for the water wave problem. In other words, we illustrate the advantage of using a variational principle involving as many dependent variables as possible. We emphasize that our primary purpose here is to provide a generalized framework for deriving model equations for water waves. This methodology is explained on various examples, some of them being new to our knowledge. However, the potential of the present approach is far from being fully exploited.

The paper is organized as follows. In Section~\ref{secrellag}, Luke's Lagrangian is {\em relaxed\/} to incorporate explicitly more degrees of freedom. This modification yields the Hamilton principle in its most general form. The advantage of this formulation is subsequently illustrated with examples over a fixed horizontal bottom, for the sake of simplicity. We begin in Section~\ref{secsw} with shallow water models, where some well-known and a few novel models are derived from the same Lagrangian, but with different subordinate constraints. In Section~\ref{secdw}, similar model equations are derived in the deep water limit. In particular, a generalization of the Klein--Gordon equation and a remarkably simple accurate approximation for traveling waves are derived. A generalized ansatz, including the shallow and deep waters as limiting particular cases is presented in Section~\ref{secfd}. Further generalizations are discussed in Section~\ref{secgenans} and their advantage is illustrated with a variant of Serre's equations. Finally, conclusions and perspectives for future studies are outlined in Section~\ref{concl}.

%%% ------------------------------------------------------------------------ %%%

\section{Generalized variational formulation}\label{secrellag}

Water wave problem possesses several variational structures \cite{Whitham1965, Luke1967, Zakharov1968}. In the present study we will extensively exploit the Lagrangian variational formalism. Surface gravity wave equations, for a potential flow with an impermeable bottom and an impermeable free surface where the pressure is constant (taken to be zero), can be derived from the ``stationary point'' (point where the variation is zero) of the following functional \cite{Luke1967}:
\begin{equation}\label{defL0}
  \mathcal{L}\,=\int_{t_1}^{t_2}\!\int_{\Omega}\mathscr{L}\,\rho\ \ud^2\/\vx\,\ud\/t, \quad
  \mathscr{L}\,=\,-\int_{-d}^\eta\left[\,g\/y\,+\,\phi_t\,+\,\half\/(\grad\,\phi)^2\,+\,\half\/\phi_y^{\,2}\,\right]\ud\/y,
\end{equation}
with $\vx = (x_1, x_2)$ the horizontal Car\-tesian co\-ordinates, $y$ the upward vertical co\-ordinate, $t$ the time, $\grad$ the horizontal gradient, $\Omega$ the horizontal domain, $\phi$ the velocity potential and $g>0$ the acceleration due to gravity; $y = \eta(\vx, t)$, $y = 0$ and $y = -d(\vx,t)$ being, respectively, the equations of the free surface, of the still water level and of the bottom. For the sake of simplicity, the surface tension effect is neglected and the fluid density $\rho$ is constant (and can thus be set to unity without any loss of generality), but this is not a limitation for the purpose of this paper.

Integrating by parts, then neglecting the terms at the horizontal and temporal boundaries because they do not contribute to the functional variations (this will be done repeatedly below without explicit mention), Luke's variational formulation (\ref{defL0}) can be rewritten with the following Lagrangian density:
\begin{equation}\label{defL1}
  \mathscr{L}\, =\,\phis\,\eta_t\,+\,\phib\,d_t\,-\,{\half}\/g\/\eta^2\,+\,\half\/g\/d^2\,-\,\int_{-d}^{\,\eta}\left[\,\half\/(\nab\,\phi)^2\,+\,\half\/\phi_y^{\,2}\,\right]\ud\/y,
\end{equation}
where the over `tildes' and `wedges' denote, respectively, the quantities written at the free surface $y = \eta$ and at the bottom $y = -d$. We shall also denote with `bars' the quantities averaged over the water depth, \eg
\begin{equation*}
  \vum(\vx,t)\ \equiv\ \frac{1}{\eta(\vx,t)+d(\vx,t)}\int_{-d(\vx,t)}^{\eta(\vx,t)}\,\vu(\vx,y,t)\,\ud\/y.
\end{equation*}

The variational formulations (\ref{defL0}) and (\ref{defL1}) impose that any approximation is exactly irrotational, \ie, the choice of an ansatz for $\phi$ necessarily implies an irrotational motion. Note that the term $\half gd^2$ in (\ref{defL1}) can be omitted because, $d$ being prescribed, it does not contribute to the variation process. Note also that Luke's Lagrangian \eqref{defL0} can be extended to the case where the bottom function $d(\vx,t)$ is unknown if a condition at $y = -d$ is added to the problem \cite{Thomas1988}.

To give us more freedom while keeping an exact formulation, the variational principle is modified (relaxed) by introducing explicitly the horizontal velocity $\vu = \nab\phi$ and the vertical one $v = \phi_y$. The variational formulation can thus be reformulated with the Lagrangian density
\begin{equation}\label{defL2ori}
  \mathscr{L} = \phis\eta_t + \phib d_t-\half\/g\eta^2-\int_{-d}^{\eta}\left[\half(\vu^2+v^2)+\vmu\scal(\nab\phi - \vu) + \nu(\phi_y-v)\right]\ud\/y,
\end{equation}
where the Lagrange multipliers $\vmu$ and $\nu$ have to be determined. By variations with respect of $\vu$ and $v$, one finds at once the definition of the Lagrange multipliers: 
\begin{equation*}
  \vmu\ =\ \vu, \qquad \nu\ =\ v. 
\end{equation*}
Using these definitions, (\ref{defL2ori}) becomes
\begin{equation}\label{defL2}
  \mathscr{L}\, =\,\phis\,\eta_t\,+\,\phib\,d_t\,-\,\half\,g\,\eta^2\,+\,\int_{-d}^{\,\eta}\left[\,\half\,\vu^2\,+\,\half\,v^2\,-\,\vu\scal\nab\phi\,-\,v\,\phi_y\,\right]\ud\/y.
\end{equation}
However, it is advantageous to keep the most general form of the Lagrangian \eqref{defL2ori}, because it allows to choose ansatz for Lagrange multipliers $\vmu$ and $\nu$ that are different from the velocity field $\vu$ and $v$. Indeed, the Lagrangian density (\ref{defL2ori}) involves six variables \{$\eta, \phi, \vu, v, \vmu, \nu$\}, while the simplified Lagrangian (\ref{defL2}) involves only four variables \{$\eta, \phi, \vu, v$\} and the original Lagrangian (\ref{defL1}) only two ($\eta$ and $\phi$). These additional variables introduce additional freedom in the construction of approximations, thus allowing more subordinate relations to be fulfilled. The Lagrangian density (\ref{defL2}) was used by \textsc{Kim} {\em et al.} \cite{Kim2001} to derive the `irrotational' Green--Naghdi equations for long waves in shallow water. The more general Lagrangian density (\ref{defL2ori}) provides more flexibility to derive model equations, as illustrated below.

The connection of (\ref{defL2ori}) with the variational formulation of the classical mechanics can be seen applying Green's theorem to (\ref{defL2ori}) that yields another equivalent variational formulation involving the Lagrangian density
\begin{align}\label{defL3}
  \mathscr{L}\ =&\ (\eta_t+\vmus\scal\nab\eta-\nus)\,\phis\ +\ (d_t+\vmub\scal\nab d+\nub)\,\phib\ -\ \half\,g\,\eta^2\  \nonumber \\ 
  &+\ \int_{-d}^{\,\eta}\left[\,\vmu\scal\vu-{\half}\/\vu^2\,+\,\nu\/v-\half\/v^2\, +\,(\nab\scal\vmu+\nu_y)\,\phi\,\right]\ud\/y,
\end{align}
which in deep water limit ($d\rightarrow\infty$) becomes (if $\phi$ decays faster than $y^{-1}$ as $y\rightarrow-\infty$)
\begin{align*}
  \mathscr{L}\ =&\ (\eta_t+\vmus\scal\nab\eta-\nus)\,\phis\,-\,\half\,g\,\eta^2\nonumber\\ &+\, \int_{-\infty}^{\,\eta} \left[\,\vmu\scal\vu-{\half}\/\vu^2\,+\,\nu\/v-\half\/v^2\,+\, (\nab\scal\vmu+\nu_y)\,\phi\,\right]\ud\/y.
\end{align*} 
Thus, in the special case $\vmu = \vu$ and $\nu = v$, the Hamilton principle of classical mechanics is recovered in Eulerian description of motion, \ie, the variational principle involves the kinetic energy minus the potential energy plus some constraints for the fluid incompressibility, for the flow irrotationality and for the bottom and surface impermeabilities, as already pointed out by \textsc{Miles} \cite{Miles1977}. In other words, the Lagrangian density (\ref{defL3}) is the Hamilton principle in its most general form for irrotational surface gravity waves.

Note that, via the integration by parts, the term $\eta_t\/\phis$, for example, can be replaced by $-\eta\/\phis_t$ in all the Lagrangian densities given above, without loss (nor gain) of generality. Note also that the relaxed variational formulations involving (\ref{defL2ori}) and (\ref{defL3}) being strictly equivalent, one should use the more convenient one depending on the problem under consideration. Note finally that extensions of (\ref{defL2ori}) and (\ref{defL3}) including, \eg, obstacles, surface tensions and stratifications in several homogeneous layers are straightforward generalizations. For instance, to include the surface tension it is sufficient to add the term $-\sigma (\sqrt{1+(\nab\eta)^2} - 1)$ into the definition of the Lagrangian density (\ref{defL3}), $\sigma$ being the surface tension coefficient.

The goal in this paper is to illustrate the power of the relaxed variational principle via some simple examples. We shall thus consider, for simplicity, a fixed horizontal bottom and we shall derive various approximate equations for shallow and deep waters. Possibilities for arbitrary depths and some other generalizations will also be discussed.

%%% ------------------------------------------------------------------------ %%%

\section{Examples in shallow water}\label{secsw}

Let consider first the shallow water case, with constant depth for the clarity of exposition. We introduce a realistic ansatz for these waves and then apply several constraints to derive various approximations, some of them being well-known, others being new.

\subsection{Choice of a simple ansatz}

For a long wave in shallow water, in potential motion on a horizontal impermeable sea bed at $y = -d$, it has long been noticed that the velocity field can be well approximated truncating the following expansion due to \textsc{Lagrange} \cite{Lagrange1781}:
\begin{equation}\label{eq:swe}
  \vu\ =\ \vub\ -\ \half(y+d)^2\, \nab^2\vub\ +\ {\textstyle{1\over24}}(y+d)^4\,\nab^4\vub\ +\ \cdots.
\end{equation}
All Lagrange's followers (\eg, \textsc{Airy}, \textsc{Boussinesq}, \textsc{Rayleigh} and many others) used this type of expansions to derive their respective approximations \cite{Craik2004}. Reviews on shallow water approximations can be found in \textsc{Bona} {\em et al.\/} \cite{BCS, Bona2004}, \textsc{Kirby} \cite{Kirby1997}, \textsc{Madsen} and \textsc{Sch\"affer} \cite{Madsen1999}, \textsc{Wu} \cite{Wu2001a}, \textsc{Dougalis} \& \textsc{Mitsotakis} \cite{DMII}, among others.

We consider here a simple ansatz of polynomial type, that is a zeroth-order polynomial in $y$ for $\phi$ and for $\vu$, and a first-order one for $v$, \ie, we approximate flows that are nearly uniform along the vertical direction. Our ansatz thus reads
\begin{equation}\label{anssha}
  \phi\ \approx\ \phim(\vx,t), \qquad \vu\ \approx\ \vum(\vx,t), \qquad v\ \approx\ (y+d)\,(\eta+d)^{-1}\ \vs(\vx,t).
\end{equation}
Such ansatz are the basis of most shallow water approximations. We have also to introduce suitable ansatz for the Lagrange multiplier $\vmu$ and $\nu$. Since $\vmu = \vu$ and $\nu = v$ for the exact solution, a natural ansatz for the multipliers is
\begin{equation}\label{ansshamunu}
  \vmu\ \approx\ \vmum(\vx,t), \qquad \nu\ \approx\ (y+d)\,(\eta+d)^{-1}\ \nus(\vx,t).
\end{equation}
With the ansatz (\ref{anssha}) and (\ref{ansshamunu}), the Lagrangian density (\ref{defL3}) becomes
\begin{eqnarray}\label{defL3sw}
  \mathscr{L}\ &=&\ (\eta_t+\vmum\scal\nab\eta)\,\phim\ -\ \half\,g\,\eta^2\  \nonumber \\
  &&+\ (\eta+d)\left[\,\vmum\scal\vum\ -\ {\half}\,\vum^2\ +\ \third\,\nus\,\vs\ -\ \sixth\,\vs^2\, +\ \phim\,\nab\scal\vmum\,\right].
\end{eqnarray}
Using the Green formula, the variational problem can also be written such that the Lagrangian density is in the following simpler form
\begin{equation}\label{defL3swbis}
  \mathscr{L}\ =\ \phim\,\eta_t\ -\ \half\,g\eta^2\  +\ (\eta+d)\left[\,\vmum\scal\vum\, -\, {\half}\vum^2\, +\, \third\,\nus\vs\, -\, \sixth\vs^2\, -\, \vmum\scal\nab\phim\,\right]. \qquad
\end{equation}
The two Lagrangian densities (\ref{defL3sw}) and (\ref{defL3swbis}) differing by a divergence term, they yield exactly the same equations. Thus, depending on the constraints, we use the Lagrangian density leading to the simpler expression. We now investigate equations led by this shallow water model under various subordinate relations.

\subsection{Unconstrained approximation}\label{secswirr}

Without further constraints, the Euler--Lagrange of (\ref{defL3swbis}) yield 
\begin{eqnarray} \label{minL3sw1}
  \udelta\,\vum\/:&& \vzero\ =\ \vmum\ \ -\ \vum, \label{minL3sw1e1} \\
  \udelta\,\vs\/:&& 0\ =\ \nus\ - \ \vs,  \label{minL3sw1e2} \\
  \udelta\,\vmum\/:&& \vzero\ =\ \vum\ \ -\ \nab\phim, \label{minL3sw1e3} \\
  \udelta\,\nus\/:&& 0\ =\ \vs,  \label{minL3sw1e4} \\
  \udelta\,\phim\/:&& 0\ =\ \eta_t\ +\ \nab\scal\left[\,(\eta+d)\,\vmum\,\right], \label{minL3sw1e5} \\
  \udelta\,\eta\/:&& 0\ =\ \vmum\scal\vum\ -\ {\half}\,\vum^2\ +\ \third\,\nus\,\vs\ -\ \sixth\,\vs^2\ -\ \vmum\scal\nab\phim\ -\ \phim_t\ -\ g\,\eta.\label{minL3sw1e6}
\end{eqnarray} 
The relations (\ref{minL3sw1})--(\ref{minL3sw1e4}) imply that the motion is exactly irrotational, but the fluid incompressibility is not satisfied identically. With these four relations, the last two equations can be rewritten in the form:
\begin{eqnarray} \label{eqswsv}
  h_t\ +\ \nab\scal[\,h\,\vum\,]\ &=&\ 0,  \\
  \vum_t\ +\ \half\,\nab\,\vum^2\ +\ g\,\nab\,h\ &=&\ \vzero,\label{eqswsv2}
\end{eqnarray}
where $h=\eta+d$ is the total water depth. Equations (\ref{eqswsv})--(\ref{eqswsv2}) are the very well-known nonlinear shallow water equations, also known as Airy or Saint-Venant equations (\textsc{Wehausen} \& \textsc{Laitone} \cite{Wehausen1960}, \textsection 28). They are sometimes called {\em non-dispersive fully-nonlinear approximation} because their classical derivation assumes long waves without the extra hypothesis of small amplitudes. These equations have a canonical Hamiltonian structure for the conjugate variables $\eta$ and $\phim$ with the Hamiltonian
\begin{equation*}
  \int_{\Omega}\left\{\,\half\,g\,\eta^2\,+\,\half\,(\eta+d)\left(\nab\phim\right)^2\,\right\}\,\ud^2\/\vx.
\end{equation*}

The Saint-Venant equations do not admit smooth progressive wave solutions. They are nevertheless widely used because they can be solved analytically by the method of characteristics \cite{Stoker1957}. Moreover, numerous efficient finite volumes type schemes have been proposed \cite{Zhou2002}. These equations have also been validated by comparisons with experimental data even for the wave runup case \cite{Synolakis1987}. Below we present several more elaborated models.

\subsection{Constraining with free surface impermeability}
\label{secswinc1}

We now constrain the ansatz (\ref{anssha}) imposing that the impermeability of the free surface is satisfied identically. Since the surface impermeability is expressed through the velocity $(\vmu,\nu)$ in (\ref{defL3}), we substitute
\begin{equation}\label{consvsw}
  \nus\ =\ \eta_t\ +\ \vmum\scal\nab\eta,
\end{equation}
into the Lagrangian density (\ref{defL3sw}), and the subsequent Euler--Lagrange equations are
\begin{eqnarray} \label{minL3sw2}
  \udelta\,\vum\/:&& \vzero\ =\ \vmum\ \ -\ \vum, \label{minL3sw2e1} \\
  \udelta\,\vs\/:&& 0\ =\ \eta_t\ +\ \vmum\scal\nab\eta\ - \ \vs, \label{minL3sw2e2}\\
  \udelta\,\vmum\/:&& \vzero\ =\ \vum\ +\ \third\,\vs\,\nab\eta\ -\ \nab\phim, \label{minL3sw2e3} \\
  \udelta\,\phim\/:&& 0\ =\ \eta_t\ +\ \nab\scal\left[\,(\eta+d)\,\vmum\,\right], \label{minL3sw2e4} \\
  \udelta\,\eta\/:&& 0\ =\ \vmum\scal\vum\ -\ {\half}\,\vum^2\ -\ \sixth\,\vs^2\ -\ \vmum\scal\nab\phim\ -\ \phim_t\ -\ g\,\eta\nonumber\\
  &&\qquad\ -\ \third\,(\eta+d)\left[\,\vs_t\,+\,\vmum\scal\nab\vs\,+\,\vs\,\nab\scal\vmum\,\right].\label{minL3sw2e5}
\end{eqnarray}
The relations (\ref{minL3sw2}) and (\ref{minL3sw2e3}) link the velocity potential and the horizontal velocity as $\nab\phim\neq\vum = \vmum$ and, therefore, equations (\ref{minL3sw2})--(\ref{minL3sw2e5}) cannot be derived from Luke's variational principle. Relations (\ref{minL3sw2}) and (\ref{minL3sw2e4}) provide the mass conservation and hence, with (\ref{consvsw}), the approximation (\ref{minL3sw2})--(\ref{minL3sw2e5}) implies that the fluid incompressibility is fulfilled identically.

Eliminating $\phim$, $\vmum$ and $\vs$ from the horizontal gradient of (\ref{minL3sw2e5}), the system (\ref{minL3sw2})--(\ref{minL3sw2e5}) becomes
\begin{align}\label{eqswser}
  h_t + \nab\scal[h\vum] &= 0, \\
  \qquad \vum_t + \half\nab\vum^2 + g\nab h + \third h^{-1}\nab[h^2\tilde{\gamma}] &= \third(\vum\scal\nab h)\/\nab(h\nab\scal\vum) - \third\left[\/\vum\scal\nab(h\nab\scal\vum)\/\right]\nab h,\label{eqswser2}
\end{align}
with $h = \eta+d$ and where
\begin{equation}\label{eqgamma_tilde}
  \tilde{\gamma}\ =\ \vs_t\ +\ \vum\scal\nab\vs \ =\ h\ \left\{\ (\nab\scal\vum)^2\,-\,\nab\scal\vum_t\,-\,\vum\scal\nab\,[\,\nab\scal\vum\,]\ \right\},
\end{equation}
is the fluid vertical acceleration at the free surface.

In the two-dimensional case (one horizontal dimension) the right-hand side of (\ref{eqswser2}) vanishes and the system (\ref{eqswser}), (\ref{eqswser2}) reduces to the equations first derived by \textsc{Serre} \cite{Serre1953}, independently rediscovered by \textsc{Su} and \textsc{Gardner} \cite{Su1969} and again by \textsc{Green}, \textsc{Laws} and \textsc{Naghdi} \cite{Green1974}. It is sometimes called {\em weakly-dispersive fully-nonlinear approximation} \cite{Wu2001a}. These equations admit a traveling solitary wave solution
\begin{equation*}
  \eta\ =\ a\ \mathrm{sech}^2\,\half\/\kappa\/(\/x_1-c\/t\/), \qquad c^2\ =\ g\,(\/d+a\/), \qquad (\/\kappa\/d\/)^2\ =\ 3\,a\,(d+a)^{-1},
\end{equation*}
which is linearly stable \cite{Li2002}. Note that this solution does not impose any limitation on the wave amplitude, meaning that Serre's equations are physically inconsistent for the highest waves. Note also that the Serre equations have a {\em non-canonical} Hamiltonian structure \cite{Li2002}.

In three dimensions, equations (\ref{eqswser})--(\ref{eqswser2}) were called by \textsc{Kim} \etal \cite{Kim2001} `irrotational' Green--Naghdi equations. If the right-hand side of (\ref{eqswser2}) is neglected, we recover the classical Green--Naghdi equations \cite{Green1976}.

\textsc{Craig} and \textsc{Grooves} \cite{Craig1994}, and many others, also derived various shallow water models starting from a variational principle. The main difference with our approach is that we do not introduce any small parameter but we make assumptions on the vertical structure of the flow.

\subsection{Constraining with incompressibility and partial potential flow I}\label{secswincpot}

Here, we restrict the freedom imposing that the velocity potential is related to the horizontal velocity as $\vu = \nab\phi$, and that the fluid incompressibility $\nab\scal\vu + v_y = 0$ is fulfilled, together with the relations $\vmu = \vu$ and $\nu = v$, \ie, we take the subordinate conditions
\begin{equation*}
  \vmum\ =\ \vum, \qquad \nus\ =\ \vs, \qquad \vum\ =\ \nab\phim, \qquad \vs\ =\ -(\eta+d)\,\nab^2\phim.
\end{equation*}
These constraints do not impose exact irrotationality because $v\neq\phi_y$. Obviously, we shall derive an approximation which lies ``between'' the Saint-Venant and Serre equations.

Thus, the Lagrangian density (\ref{defL3swbis}) becomes
\begin{equation}\label{LaggKB}
  \mathscr{L}\ =\ \phim\,\eta_t\ -\ \half\,g\,\eta^2\ -\ \half\,(\eta+d)\left(\nab\phim\right)^2 \ +\ {\sixth}\,(\eta+d)^3\left(\nab^2\phim\right)^2,
\end{equation}
and its Euler--Lagrange equations yield
\begin{eqnarray*}
  \udelta\,\phim\/:&& 0\ =\ \eta_t\ +\ \nab\scal\left[\,(\eta+d)\,\nab\phim\,\right]\ +\ \third\,\nab^2\left[\,(\eta+d)^3\left(\nab^2\phim\right)\,\right], \label{minL3sw3e1} \\
  \udelta\,\eta\/:&& 0\ =\  \phim_t\ +\ g\,\eta\ +\ \half\,\left(\nab\phim\right)^2 \ -\ \half\,(\eta+d)^2\left(\nab^2\phim\right)^2.
\end{eqnarray*}
It seems that these equations have never appeared before in the literature. They are a generalization of the so-called Kaup--Boussinesq (or canonical Boussinesq) equations \cite{Kaup1975, Kupershmidt1985} and are thus referred to as the gKB equations. This can be seen noticing that the gKB equations can be derived from the canonical Hamiltonian
\begin{equation}\label{Hkaup}
  \int_{\Omega}\left\{\,\half\,g\,\eta^2\,+\,\half\,(\eta+d)\left(\nab\phim\right)^2\, -\,{\sixth}\,(\eta+d)^3 \left(\nab^2\phim\right)^2\,\right\}\,\ud^2\/\vx,
\end{equation}
while the classical Kaup--Boussinesq (cKB) equations are obtained replacing $(\eta+d)^3$ by $d^3$ in (\ref{Hkaup}) and restricting the resulting Hamiltonian to one horizontal dimension. Note that the Lagrangian (\ref{LaggKB}) is cubic in $\eta$, meaning that it has (at most) one local minimum in $\eta$ and not a global one.

The linearized gKB and cKB systems admit the special traveling wave solution
\begin{equation}
  \eta\ =\ a\cos k(x_1-ct), \qquad c^2\ =\ g\/d\,(1-\third\/k^2\/d^2),  \label{disrelsw3}
\end{equation}
implying that these equations are linearly ill-conditioned ($c^2 < 0$ for $k d > \sqrt{3}$). However, if, like the cKB, the gKB equations are integrable, they may be a somewhat interesting model for gravity waves in shallow water.

\subsection{Constraining with incompressibility and partial potential flow II}\label{secswincpot2}

So far, all the approximations derived turned out to be such that $\vmu = \vu$ and $\nu = v$. We propose here a novel approximation that does not satisfies one of these identities and that is an interesting variant of the previous model.

We impose a partially potential flow such that $\vmu = \nab\phi$ and $\nu = \phi_y$, together with the incompressibility condition $\nab\scal\vu + v_y = 0$ and the condition $\vu = \vmu$. Thus, substituting the constraints
\begin{equation*}
  \vmum\ =\ \vum\ =\ \nab\phim, \qquad \nus\ =\ 0,  \qquad \vs\ =\ -(\eta+d)\,\nab^2\phim,
\end{equation*}
into the Lagrangian density (\ref{defL3swbis}) yields
\begin{equation*} 
  \mathscr{L}\ =\ \phim\,\eta_t\ -\ \half\,g\,\eta^2\ -\ \half\,(\eta+d)\left(\nab\phim\right)^2 \ -\ {\sixth}\,(\eta+d)^3\left(\nab^2\phim\right)^2,
\end{equation*}
and the corresponding Euler--Lagrange equations are
\begin{eqnarray}\label{minL3sw4}
  \udelta\,\phim\/:&& 0\ =\ \eta_t\ +\ \nab\scal\left[\,(\eta+d)\,\nab\phim\,\right]\ -\ \third\,\nab^2\left[\,(\eta+d)^3\left(\nab^2\phim\right)\,\right], \label{minL3sw4e1} \\
  \udelta\,\eta\/:&& 0\ =\  \phim_t\ +\ g\,\eta\ +\ \half\,\left(\nab\phim\right)^2 \ +\ \half\,(\eta+d)^2\left(\nab^2\phim\right)^2. \label{minL3sw4e2}
\end{eqnarray}
These equations can be derived from the canonical Hamiltonian
\begin{equation*}
  \int_{\Omega}\left\{\,\half\,g\,\eta^2\,+\,\half\,(\eta+d)\left(\nab\phim\right)^2\, +\,{\sixth}\,(\eta+d)^3 \left(\nab^2\phim\right)^2\,\right\}\,\ud^2\/\vx,
\end{equation*}
which is always positive (an interesting feature for modeling water waves). To the linear approximation, equations (\ref{minL3sw4}), (\ref{minL3sw4e2}) have the progressive wave solution
\begin{equation}
  \eta\ =\ a\cos k(x_1-ct), \qquad c^2\ =\ g\/d\,(1+\third\/k^2\/d^2),\label{disrelsw4}
\end{equation}
which is well-behaved (\ie, $c^2$ is never negative). Comparisons with the gKB equations suggest to refer to equations (\ref{minL3sw4})--(\ref{minL3sw4e2}) as {\em regularized general Kaup-Boussinesq} (rgKB). However, the linear dispersion relation (\ref{disrelsw4}) approximates the dispersion relation of linear waves --- \ie, $c^2=g\tanh(k d)/k$ --- only to the order $\operatorname{O}(k^2)$, while (\ref{disrelsw3}) is $\operatorname{O}(k^4)$. Therefore, the rgKB equations are not very interesting for modeling water waves, but these equations may be of interest to model other physical processes.

\subsection{Other constraints and generalizations}\label{secswrem}

We briefly discuss now some other possible generalizations.

\subsubsection{Constraining with incompressibility and potential flow I}
\label{secswincpot3}

In the previous example, we have constructed an approximation such that $\vmu = \vu$ but $\nu \neq v$. Now, we release the constraint $\vmu = \vu$ and keep the other constraints. Thus, we impose
\begin{equation*}
  \vmum\ =\ \nab\phim, \qquad \nus\ =\ 0,  \qquad \vs\ =\ -(\eta+d)\,\nab\scal\vum,
\end{equation*}
so that the pseudo velocity field ($\vmu,\nu$) is irrotational while the velocity field ($\vu,v$) is incompressible. After same elementary algebra, the Lagrangian density becomes
\begin{equation*}
  \mathscr{L}\ =\ \phim\,h_t\ -\ \half\,g\,\eta^2\ +\ h\,\vum\scal\nab\phim\ -\ \half\,h\,\vum^2\ -\ {\sixth}\,h^3\left(\nab\scal\vum\right)^2\ -\ h\left(\nab\phim\right)^2,
\end{equation*}
where $h = \eta + d$. The Euler--Lagrange equations give
\begin{eqnarray*}
  \udelta\,\vum\/:&& \vzero\ =\ h\,\nab\phim\ -\ h\,\vum\ +\ \third\,\nab\left[\,h^3\,\nab\scal\vum\,\right], \label{minL3sw5e1} \\
  \udelta\,\phim\/:&& 0\ =\ h_t\ -\ \nab\scal\left[\,h\,\vum\,\right]\ +\ 2\,\nab\scal\left[\,h\,\nab\phim\,\right], \label{minL3sw5e2} \\
  \udelta\,\eta\/:&& 0\ =\  \phim_t\ +\ g\,\eta\ +\ \half\,\vum^2\ +\,\left(\nab\phim\right)^2\ -\ \vum\scal\nab\phim \ +\ \half\,h^2\left(\nab\scal\vum\right)^2. \label{minL3sw5e3}
\end{eqnarray*}
The linearization of this system of equations have a $(2\pi/k)$-periodic sinusoidal traveling wave solution with the dispersion relation
\begin{equation*}
  c^2\ =\ g\/d\,(1+\twothird\/k^2\/d^2)\,(1+\third\/k^2\/d^2)^{-1}\ =\ g\/d\,(1+\third\/k^2\/d^2)\,+\,\operatorname{O}(k^4),
\end{equation*}
which, like the previous example, is not satisfactory for water waves. However, these equations may be of interest in other contexts than water waves.

\subsubsection{Constraining with incompressibility and potential flow II}
\label{secswincpot4}

We now assume that the pseudo velocity field ($\vmu, \nu$) is divergence free, while the velocity field ($\vu, v$) is irrotational, \ie, we impose the constraints  
\begin{equation*}
  \vum\ =\ \nab\phim, \qquad \vs\ =\ 0,  \qquad \nus\ =\ -(\eta+d)\,\nab\scal\vmum.
\end{equation*}
The Lagrangian density becomes
\begin{equation*}
  \mathscr{L}\ =\ \phim\,\eta_t\ -\ \half\,g\,\eta^2\  -\ \half\,(\eta+d)\/\left(\nab\phim\right)^2,
\end{equation*}
which yields the Saint-Venant equations. Thus, these constraints do not bring anything new. It should be emphasized that this is the case for the special shallow water ansatz we are considering here, but this is not necessarily the case for other ansatz.

\subsubsection{Further possibilities}

The constraints of Sections~\ref{secswincpot} to \ref{secswincpot4} can be unified into a single formalism considering combinations. Indeed, the velocity field ($\vu,v$) being not more (nor less) physical than the pseudo-velocity field ($\vmu,\nu$) and the potential velocity field ($\nab\phi, \phi_y$), the constraints can be imposed by combinations of these three fields. For instance, we could impose the irrotationality for the field
\[
  (\,c_1\/\vu\, +\, c_2\/\vmu\, +\, (1-c_1-c_2)\/\nab\phi\ ,\ c_1\/v\, +\,c_2\/\nu\,+\,(1-c_1-c_2)\/\phi_y\,),
\]
the fluid incompressibility for the field
\[
  (\,c_3\/\vu\, +\, c_4\/\vmu\, +\, (1-c_3-c_4)\/\nab\phi\ ,\ c_3\/v\, +\,c_4\/\nu\,+\,(1-c_3-c_4)\/\phi_y\,),
\]
and so on for any constraint we may think of. The $c_n$ are parameters at our disposal. We can choose them in a convenient way based on some mathematical and physical considerations. For example, imposing that the approximate equations derived must be linearly well-posed and/or have better dispersion relation 
properties.

In the examples above, only some kinematic constraints (irrotationality, incompressibility, impermeability) were used. We could have also considered dynamical constraints based on, \eg, the Bernoulli equation, or other relevant dynamical equations.

The relaxed variational principle provided a common platform for deriving several shallow water equations from the same ansatz in changing only the constraints. Beside the ansatz, no further approximation were made and the derivations required only some elementary algebra. Using more general ansatz --- \ie, involving more free functions and parameters --- one can introduce more constraints, if desired, and derives more accurate approximations. A simple example is given in Section~\ref{secgenans} below.

%%% ------------------------------------------------------------------------ %%%

\section{Examples in deep water}
\label{secdw}

We illustrate here the advantages of the relaxed variational principle in the opposite limiting case of deep water.

\subsection{Choice of an ansatz}

For progressive waves in deep water, the Stokes expansion shows that the velocity field varies nearly exponentially along the vertical (Appendix~\ref{appsto}). Even for very large unsteady waves (including breaking waves), accurate numerical simulations and experiments have shown that the vertical variation of the velocity field is indeed very close to an exponential \cite{Grue2003, Jensen2007}. Thus, this property is exploited here to derive simple approximations for waves in deep water.

Let $\kappa > 0$ be a characteristic wavenumber corresponding, for example, to the carrier wave of a modulated wave group or to the peak frequency of a JONSWAP spectrum. Following the discussion above, it is natural to seek approximations in the form
\begin{equation}\label{ansinf}
  \{\,\phi\,;\,\vu\,;\,v\,;\,\vmu\,;\,\nu\,\}\ \approx\ \{\,\phis\,;\,\vus\,;\,\vs\,; \,\vmus\,;\,\nus\,\}\ \ue^{\kappa\/(y-\eta)},
\end{equation}
where $\phis$, $\vus$, $\vs$, $\mus$ and $\nus$ are functions of $\vx$ and $t$ that will be determined using the variational principle. The ansatz (\ref{ansinf}) is certainly the simplest possible that is consistent with experimental evidences. This ansatz has already been used by \textsc{Kraenkel} {\em et al.} \cite{Kraenkel2005} for building their approximation. Possible generalizations are discussed in Section \ref{secgenans}.

For the sake of simplicity, we introduce the constraints $\vmus = \vus$ and $\nus = \vs$. Thus, the ansatz (\ref{ansinf}) substituted into the Lagrangian density (\ref{defL2}) yields
\begin{equation}\label{defL21}
  2\/\kappa\,\mathscr{L}\ =\ 2\/\kappa\,\phis\,\eta_t\,-\,g\/\kappa\,\eta^2\,+\, \half\/\vus^2\,+\,\half\/\vs^2\,-\,\vus\scal(\nab\phis-\kappa\/\phis\nab\eta)\,-\,\kappa\,\vs\,\phis.
\end{equation}
With (or without) subordinate relations, this Lagrangian gives various equations. We investigate two cases here.

\subsection{Unconstrained approximation}
\label{secdwirr}

Without further constraints, the Euler--Lagrange equations yield
\begin{eqnarray*}
  \udelta\,\vus\/:&& \vzero\ =\ \vus\,-\,\nab\phis\,+\,\kappa\,\phis\,\nab\eta, \\
  \udelta\,\vs\/:&&  0\ =\ \vs\,-\,\kappa\,\phis, \\
  \udelta\,\phis\/:&& 0\ =\ 2\/\kappa\,\eta_t\,+\,\nab\scal\vus\,-\,\kappa\,\vs\, +\,\kappa\,\vus\scal\nab\eta, \\
  \udelta\,\eta\/:&& 0\ =\ 2\/g\/\kappa\,\eta\,+\,2\/\kappa\,\phis_t\,+\,\kappa\,\nab\scal(\,\phis\,\vus\,).
\end{eqnarray*}
The two first relations imply that this approximation is exactly irrotational and their use in the last two equations gives
\begin{eqnarray} \label{gkg}
  & \eta_t\,+\,\half\/\kappa^{-1}\nab^2\phis\,-\,\half\/\kappa\/\phis\ =\ \half\,\phis\left[\,\nab^2\eta\,+\,\kappa\,(\nab\eta)^2\,\right],& \label{surimpL21}\\
  & \phis_t\,+\,g\,\eta\ =\ -\half\,\nab\scal\left[\,\phis\,\nab\phis\,-\,\kappa\,\phis^2\,
\nab\eta\,\right].& \label{surisoL21}
\end{eqnarray}
Since equations (\ref{surimpL21})--(\ref{surisoL21}) derive from an irrotational motion, they can also be obtained from Luke's Lagrangian (\ref{defL1}) under ansatz (\ref{ansinf}). Equations (\ref{surimpL21})--(\ref{surisoL21}) are a deep water counterpart of Saint-Venant equations for shallow water waves; this claim will appear clearer in Section \ref{secfd}. They can also be derived from the canonical Hamiltonian
\begin{equation}\label{Hgkg}
  \int_{\Omega}\left\{\,\half\,g\,\eta^2\,+\,\quat\,\kappa^{-1}\left[\,\nab\phis\,-\,\kappa\,\phis\,\nab\eta\,\right]^2\,+\,\quat\,\kappa\,{\phis}^2\,\right\}\,\ud^2\/\vx.
\end{equation}
This `simple' Hamiltonian is quartic in nonlinearities and involves only first-order derivatives. It has to be compared with Zakharov's quartic Hamiltonian (\ref{Hzak}) which involves second-order derivatives and pseudo-differential operators. However, Zakharov's quartic Hamiltonian is valid for broad spectra. Note that the Hamiltonian (\ref{Hgkg}) cannot be derived from the exact one (\ref{Hexa}), since the latter assumes that irrotationality and incompressibility are both satisfied identically in the bulk, while the incompressibility is not fulfilled by equations (\ref{gkg})--(\ref{surisoL21}).

To the linear approximation, after elimination of $\phis$, equations (\ref{gkg})--(\ref{surisoL21}) yield
\begin{equation}\label{gkglin}
  \eta_{tt}\ -\ \half\,(g/\kappa)\,\nab^2\eta\ +\ \half\,g\,\kappa\,\eta\ =\ 0,
\end{equation}
that is a Klein--Gordon equation. For this reason, equations (\ref{gkg})--(\ref{surisoL21}) will be referred here as {\em generalized Klein--Gordon\/} (gKG). The Klein--Gordon equation is prominent in mathematical physics and appears, \eg, as a relativistic generalization of the Schr\"odinger equation. The Klein--Gordon equation (\ref{gkglin}) admits a special ($2\pi/k$)-periodic traveling wave solution
\begin{equation*}
  \eta\ =\ a\,\cos k\/(x_1-c\/t), \qquad c^2\ =\ \half\,g\,(k^2+\kappa^2)\,(\kappa\,k^2)^{-1}.
\end{equation*}
Therefore, if $k = \kappa$ the exact dispersion relation of linear waves (\ie, $c^2 = g/k$) is recovered, as it should be. This means, in particular, that the gKG model is valid for spectra narrow-banded around the wavenumber $\kappa$.

We focus now on ($2\pi/\kappa$)-periodic progressive waves solution of the gKG equations, \ie, we seek for solutions depending only on the variable $\theta = \kappa(x_1\!-\!c\/t)$. We were not able to find an exact analytic solution but a Stokes-like expansion gives some interesting insights. To the seventh-order, we have
\begin{eqnarray*}
\kappa\,\eta&=&\alpha\cos\theta
 \,+\,{\textstyle{1\over2}}\alpha^2\!\left(1\!+\!{\textstyle{{\bf25}\over12}}\alpha^2\!
 +\!{\textstyle{{\bf1675}\over{\bf192}}}\alpha^4\right)\cos2\theta \nonumber \\
&& +\,{\textstyle{3\over8}}\alpha^3\!\left(1\!+\!
{\textstyle{{\bf99}\over16}}\alpha^2\!+\!{\textstyle{{\bf11807}\over320}}\alpha^4\right)\cos3\theta
\,+\,{\textstyle{1\over3}}\alpha^4\!\left(1\!+\!
{\textstyle{{\bf64}\over{\bf5}}}\alpha^2\right)\cos4\theta  \\
&&+\,{\textstyle{125\over384}}\alpha^5\!\left(1\!+\!
{\textstyle{{\bf6797}\over{\bf300}}}\alpha^2\right)\cos5\theta
\,+\,{\textstyle{27\over80}}\alpha^6\cos6\theta
\,+\,{\textstyle{16807\over46080}}\alpha^7\cos7\theta\,+\,\mathrm{O}(\alpha^8), \nonumber \\
{g^{-{1\over2}}}\/\kappa^{3\over2}\,\phis&=&\alpha\left(1\!-\!{\textstyle{1\over4}}\alpha^2\!
 -\!{\textstyle{{\bf59}\over96}}\alpha^4\!-\!{\textstyle{{\bf4741}\over1536}}\alpha^6\right)\sin\theta
 \,+\,{\textstyle{1\over2}}\alpha^2\!\left(1\!+\!{\textstyle{{\bf11}\over12}}\alpha^2\!
 +\!{\textstyle{{\bf547}\over{\bf192}}}\alpha^4\right)\sin2\theta \nonumber \\
&& +\,{\textstyle{3\over8}}\alpha^3\!\left(1\!+\!
{\textstyle{{\bf163}\over{\bf48}}}\alpha^2\!+\!{\textstyle{{\bf221}\over{\bf15}}}\alpha^4\right)\sin3\theta
\,+\,{\textstyle{1\over3}}\alpha^4\!\left(1\!+\!
{\textstyle{{\bf149}\over{\bf20}}}\alpha^2\right)\sin4\theta  \\
&&+\,{\textstyle{125\over384}}\alpha^5\!\left(1\!+\!
{\textstyle{{\bf5057}\over{\bf375}}}\alpha^2\right)\sin5\theta
\,+\,{\textstyle{27\over80}}\alpha^6\sin6\theta
\,+\,{\textstyle{16807\over46080}}\alpha^7\sin7\theta\,+\,\mathrm{O}(\alpha^8), \nonumber \\
{g^{-{1\over2}}}\/\kappa^{1\over2}\,c &=&
1+{\textstyle{1\over2}}\alpha^2
+{\textstyle{1\over2}}\alpha^4+{\textstyle{{\bf899}\over384}}\alpha^6+\mathrm{O}(\alpha^8).
\end{eqnarray*}
The expansions of $\eta$ and $\phis$ match the exact Stokes wave (\cf Appendix~\ref{appsto}) up to the third-order (non-matching coefficients are displayed bold). This is not surprising since the gKG equations are cubic in nonlinearities. A bit more surprising is that the phase velocity $c$ is correct up to the fifth-order. But the most interesting is that, to the leading order, the $n$-th Fourier coefficient is (for all $n$ up to infinity)
\begin{equation}\label{fourcoef}
  \frac{n^{n-2}\ \alpha^n}{2^{n-1}\,(n\!-\!1)!},
\end{equation}
which is also the case for the exact Stokes wave (Appendix~\ref{appsto}).

In comparison, for the cubic Zakharov equations (\ref{eqcze})--(\ref{eqcze2}), the phase velocity is correct only up to the third-order and the Fourier coefficients do not verify the asymptotic behavior (\ref{fourcoef}) (see Appendix \ref{appcze}). Truncating Zakharov's Hamiltonian at the order $N+1$ in nonlinearities, the corresponding Stokes double series is correct up to the order $N$ in the expansion parameter. But none of these higher approximations have the exact asymptotic behavior (\ref{fourcoef}) for their Fourier coefficients because they involve expansions around $\eta = 0$, while the gKG does not. Zakharov's equations are valid for broad spectra, however, unlike the gKG equations.

\subsection{Constraining with the free surface impermeability}
\label{secdwinc}

In order to satisfy the free surface impermeability identically, we take
\begin{equation*}
  \vs\ =\ \eta_t\,+\,\vus\scal\nab\eta,
\end{equation*}
and the Lagrangian density (\ref{defL21}) becomes
\begin{eqnarray} \label{defL31}
  2\/\kappa\,\mathscr{L}\ =\ \phis\,(\kappa\,\eta_t+\nab\scal\vus)\,-\,g\/\kappa\,\eta^2\,+\,\half\/\vus^2\,+\, \half\/(\eta_t+\vus\scal\nab\eta)^2,
\end{eqnarray}
while the Euler--Lagrange equations yield the relations
\begin{eqnarray} \label{minL31}
  \udelta\,\vus\/:&& \vzero\ = \vus + (\eta_t+\vus\scal\nab\eta) \nab\eta - \nab\phis, \label{minL31e1} \\
  \udelta\,\phis\/:&& 0\ = \kappa\eta_t + \nab\scal\vus, \label{minL31e2} \\
  \udelta\,\eta\/:&& 0\ = 2\/g\/\kappa\eta + \kappa\phis_t + \eta_{tt} + (\vus\scal\nab\eta)_t + \nab\scal(\vus \eta_t) + \nab\scal[(\vus\scal\nab\eta)\vus].\label{minL31e3}
\end{eqnarray}
The relation (\ref{minL31e2}) implying that $\nab\scal\vu + v_y = 0$, the solution satisfies the incompressibility identically. On the other hand, the irrotationality being not verified identically, equations (\ref{minL31})--(\ref{minL31e3}) cannot be derived from Luke's variational formulation. Note that (\ref{minL31}) yields $\nab\phis = \vus + \vs\nab\eta$ that is exact for potential flows \cite{Craig1993, Fructus2005}.

As for the shallow water case, the potential $\phis$ can be eliminated from equations (\ref{minL31}), thus yielding a deep water analog of Serre's and the Green--Naghdi equations.

To the linear approximation, relations (\ref{minL31})--(\ref{minL31e3}) can be combined into a single equation for the elevation of the free surface:
\begin{equation*}
  (\nab^2-\kappa^2)\,\eta_{tt}\ +\ 2\,g\,\kappa\,\nab^2\eta\ =\ 0,
\end{equation*}
which admits the special ($2\pi/k$)-periodic solution
\begin{equation*}
  \eta\ =\ a\,\cos k\/(x_1-c\/t), \qquad c^2\ =\ 2\,g\,\kappa\,(k^2+\kappa^2)^{-1}.
\end{equation*}
Therefore, if $k = \kappa$ the exact linear approximation is recovered, as it should be. Again, this means that this model is valid for narrow-banded spectra. How narrow will be investigated now.

\subsubsection{Two-dimensional progressive waves}

We seek now exact solutions depending only on $\xi\equiv x_1-c\/t$ (two-dimensional progressive waves). Equations (\ref{minL31e1})--(\ref{minL31e2}) yield respectively
\begin{eqnarray*}
  \phis_\xi\ =\  \tilde{u}\,+\,(\tilde{u}-c)\,\eta_\xi^{\,2}, \qquad 
  \tilde{u}\ =\ c\,\kappa\,\eta\,+\,c\,K_1,
\end{eqnarray*}
where $K_1$ is an integration constant. Substituting these relations into (\ref{minL31e3}) and multiplying the result by $\eta_\xi$, after one integration and some algebra, we obtain
\begin{equation}\label{eqetainc}
  \left(\frac{\ud\,\eta}{\ud\/\xi}\right)^{\!2}\ =\ \frac{K_0^{\,2}\,-\left(\,\kappa\,\eta\,-\,K_1\,{\hat{\/\varepsilon}}^{-2}\,\right)^2}{{\hat{\/\varepsilon}}^{-2}\left(\,\kappa\,\eta\,+\,K_1\,-\,1\,\right)^2},
\end{equation}
where
\begin{equation*}
  {\hat{\/\varepsilon}}^{\/2}\ \equiv\ 2\,g\,\kappa^{-1}\,c^{-2}\ -\ 1,
\end{equation*}
and where $K_0$ is another integration constant. Assuming that $\eta_\xi=0$ at the wave crest where $\eta = a$ and at the trough where $\eta = -b$ (hence $a+b$ is the total wave height), we find 
\begin{eqnarray*}
  K_0\ =\ \half\,\kappa\,(a+b), \qquad K_1\ =\ \half\,\kappa\,(a-b)\,{\hat{\/\varepsilon}}^{\/2}.
\end{eqnarray*}
Solving equation (\ref{eqetainc}), together with the condition of zero mean free surface elevation, we obtain an exact $(2\pi/k)$-periodic solution in the parametric form \cite{Geniet2003}:
\begin{eqnarray}\label{soltro}
  k\,\xi(\tau)  = \tau\ -\ \varepsilon\,{\hat{\/\varepsilon}}^{-1}\,\sin\tau, \qquad
  k\,\eta(\tau) = \half\,\varepsilon^2\,{\hat{\/\varepsilon}}^{-1}\ +\ \varepsilon\,\cos\tau,
\end{eqnarray}
with
\begin{eqnarray*}
  \varepsilon\ =\ \half\,k\,(a+b), \qquad k\,\kappa^{-1}\ =\ {\hat{\/\varepsilon}}\,+\,\half\,(\,{\hat{\/\varepsilon}}^{-1} + {\hat{\/\varepsilon}}\,)\,\varepsilon^2,
\end{eqnarray*}
$\varepsilon$ being a wave steepness and $\tau$ being a parametric variable such that a crest is at $\tau=0$ and the closest troughs at $\tau=\pm\pi$. This remarkably simple solution describes surface waves as trochoids. Smooth surface profiles are obtained for $0\leqslant\varepsilon<{\hat{\/\varepsilon}}$, $\varepsilon\ll{\hat{\/\varepsilon}}$ corresponding to quasi-sinusoidal solutions (infinitesimal waves). For the limiting case $\varepsilon = {\hat{\/\varepsilon}}$, in the vicinity of the crest $k(a\!-\!\eta)\sim\half\varepsilon(6k\xi)^{2\over3}$ so the solution involves a sharp angle forming a $0^\circ$ inner angle (\ie, a cusp), while the exact angle should be $120^\circ$ for irrotational traveling waves.

So far, $\kappa$ is a free parameter at our disposal. We shall now investigate various choices.

\subsubsection{Simple approximation}

A `natural' choice is to take $\kappa=k$, yielding the steepness $\hat{\varepsilon} \approx 0.596$ for the limiting wave. With this peculiar choice of $\kappa$, a Stokes-like expansion of the solution of (\ref{soltro}) is
\begin{eqnarray*}
  k\,\eta&=&\alpha\cos k\/\xi\,+\,{\textstyle{1\over2}}\alpha^2\!\left(1\!+\!{\textstyle{{\bf13}\over12}}\alpha^2\!
 +\!{\textstyle{{\bf395}\over{\bf192}}}\alpha^4\right)\cos2k\/\xi \nonumber \\
  && +\,{\textstyle{3\over8}}\alpha^3\!\left(1\!+\!{\textstyle{{\bf35}\over16}}\alpha^2\!+\!{\textstyle{{\bf1727}\over320}}\alpha^4\right)\cos3k\/\xi\,+\,{\textstyle{1\over3}}\alpha^4\!\left(1\!+\!{\textstyle{{\bf33}\over{\bf10}}}\alpha^2\right)\cos4k\/\xi  \\
  &&+\,{\textstyle{125\over384}}\alpha^5\!\left(1\!+\!{\textstyle{{\bf53}\over{\bf12}}}\alpha^2\right)\cos5k\/\xi\,+\,{\textstyle{27\over80}}\alpha^6\cos6k\/\xi\,+\,{\textstyle{16807\over46080}}\alpha^7\cos7k\/\xi\,+\,\mathrm{O}(\alpha^8),  \nonumber\\
  \sqrt{k/g}\,c &=& 1+{\textstyle{1\over2}}\alpha^2+{\textstyle{1\over2}}\alpha^4+{\textstyle{{\bf611}\over384}}\alpha^6+\mathrm{O}(\alpha^8).
\end{eqnarray*}
As for the gKB approximation, the Fourier coefficients of this Stokes-like expansion satisfy the asymptotic expression (\ref{fourcoef}) and the expansion of $c$ is exact up to the fifth-order. Therefore, taking $\kappa = k$ leads to a quite accurate approximation. However, another value of $\kappa$ may lead to an even more interesting approximation.

\subsubsection{Improved parameter}\label{optkappa}

With the trochoidal solution and the relation between the parameters obtained above, the Lagrangian density (\ref{defL31}) integrated over one wavelength yields the Lagrangian
\begin{eqnarray}
  \mathcal{L}\ =\ \frac{\pi\,g\,\varepsilon^4}{4\,k^3}\ =\ \frac{\pi\,g}{k^3} \left[\rule{0mm}{7.5mm}-1\, +\,\frac{\kappa\,c^2}{2\,g}\,+\,\sqrt{\,\frac{k^2\,c^2}{2\,g\,\kappa}\,-\,\frac{k^2\,c^4}{4\,g^2}\,}\,\right]^2.
\end{eqnarray}
Thus, the Lagrangian is minimum when the steepness is minimum or, equivalently, the wavelength is maximum if the wave height is kept constant. The variational principle being defined with fixed horizontal and temporal boundaries, an improved parameter $\kappa$ can be obtained minimizing $\mathcal{L}$ keeping $k$ and $c$ constant. Thus, after some algebra, the equation $\ud\mathcal{L}/\ud\kappa = 0$ gives two possible solutions for $\kappa$:
\begin{equation}\label{kappmopt}
  \kappa_\pm\ =\ \sqrt{2}\,k\left[\,{1\,+\,2\/\varepsilon^2\,\pm\,\sqrt{1-4\varepsilon^2}}\,\right]^{-{1\over2}},
\end{equation}
provided that $\varepsilon\leqslant1/2$. Both solutions correspond to an extremum of $\mathcal{L}$. However, a solution is stable only if $\kappa$ is a minimum of $\mathcal{L}$, \ie, if
\begin{equation*}
  \left.\frac{\ud^2\,\mathcal{L}}{\ud\/\kappa^2}\right|_{\kappa=\kappa_\pm} >\ 0,
\end{equation*}
which, after some algebra, yields the condition
\begin{equation}\label{stabcri}
  \sqrt{3}\ \varepsilon\ \kappa_\pm\ <\ k.
\end{equation} 

The first solution 
\begin{equation*}
  \kappa_+\,/\,k\ =\ \sqrt{2}\left[\,{1\,+\,2\/\varepsilon^2\,+\,\sqrt{1-4\varepsilon^2}}\,\right]^{-{1\over2}}\ =\ 1\ +\ \half\,\varepsilon^4\ +\ \mathrm{O}\!\left(\varepsilon^{6}\right),
\end{equation*}
is very close to the wavenumber $k$ when the steepness $\varepsilon$ is small. When the steepness increases form $0$ to $1/2$, the dimensionless parameters $\kappa/k$, $kc^2/g$, $k^3\mathcal{L}/g$ and $\sqrt{3}\/\varepsilon\/\kappa/k$ increase monotonically. The highest wave is obtained for $\varepsilon = 1/2$, where $\kappa/k = 2/\sqrt{3}\approx1.15$, $kc^2/g = 3\sqrt{3}/4\approx1.3$ and $\sqrt{3}\/\varepsilon\/\kappa/k=1$. All these waves are smooth and should be stable because the condition (\ref{stabcri}) is fulfilled.

The second solution 
\begin{equation*}%\label{kapoptplus}
  \kappa_-\,/\,k\ =\ \sqrt{2}\left[\,{1\, + \,2\/\varepsilon^2\, - \,\sqrt{1-4\varepsilon^2}}\,\right]^{-{1\over2}},
\end{equation*}
is admissible (\ie, real) if $1/2\geqslant\varepsilon\geqslant\sqrt{\sqrt{5}-2}\approx0.486$. As the steepness decays from $1/2$ to $\sqrt{\sqrt{5}-2}$, the crest sharpens, the limiting value $\varepsilon = \hat{\varepsilon}=\sqrt{\sqrt{5}-2}$ corresponding to a cusp at the crest. Thus the sharp-crested wave is not the highest one. All these waves should be unstable because the condition (\ref{stabcri}) is violated, \ie, the solution $\kappa_-$ is a maximum of the Lagrangian and not a minimum.

\begin{figure}
  \centering
  \includegraphics[width=100mm,height=50mm]{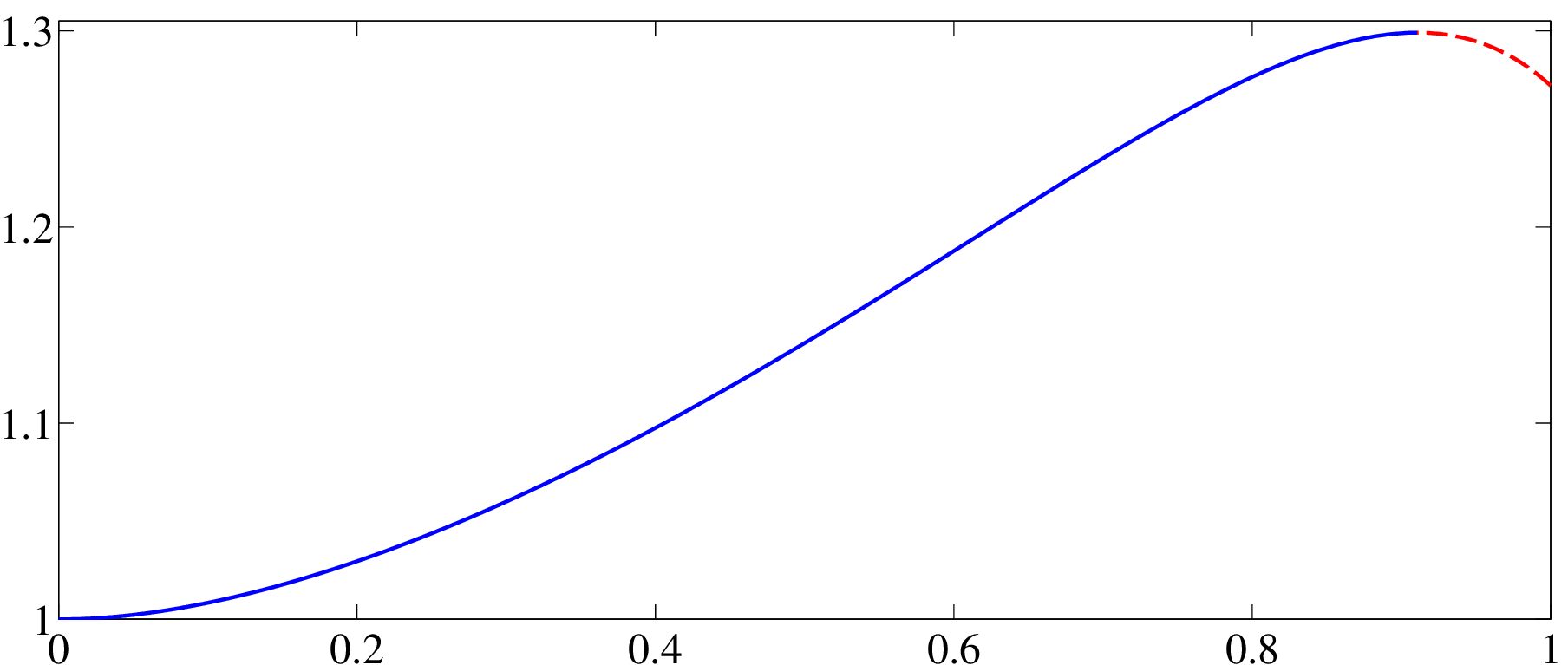}
  \put(-150,-16){\small$\tilde{u}(0)\,/\,c$}
  \put(-310,72){\Large$\frac{k\,c^2}{g}$}
  \caption{\small\em Deep water phase velocity as a function of the fluid speed at crest for the improved solution derived in \textsection \ref{optkappa}. Solid line: $\kappa_+$ solution; Dashed line: $\kappa_-$ solution.}
  \label{figc2}
\end{figure}

Note that the relations linking the parameters can be conveniently rewritten 
\[
\frac{k\,c^2}{g}\, =\, \frac{2\,\hat{\/\varepsilon}}{1\/+\/\hat{\/\varepsilon}^2}\,+\,
\frac{\varepsilon^2}{\hat{\varepsilon}}, \qquad
\hat{\/\varepsilon}\, =\, \frac{(k/\kappa_\pm)\,+\,\sqrt{\/(k/\kappa_\pm)^2-2\varepsilon^2
-\varepsilon^4\/}}{2\/+\/\varepsilon^2},
\] 
which, together with (\ref{kappmopt}), allow to express $kc^2/g$ as function of $\varepsilon$ only. Another parameter of physical interest is the fluid velocity at crest $\tilde{u}(0)$ divided by the phase velocity $c$, \ie,
\[
  \frac{\tilde{u}(0)}{c}\ =\ \frac{\varepsilon\left(\/\varepsilon\/+\/2\/\hat{\/\varepsilon}\/+\/\varepsilon\/\hat{\/\varepsilon}^2\/\right)\kappa_\pm}{2\,\hat{\/\varepsilon}\,k}.
\]
The variations of $c^2$ with respect to this parameter are displayed on Figure \ref{figc2}.

\subsubsection{Remarks}

Taking the improved parameter $\kappa$ has increased somewhat the accuracy of the approximation; the main gain is qualitative, however. Indeed, we found that waves should be unstable before a sharp crest is formed. Such behavior is not predicted by low-order perturbation expansions. This simple example is a remarkable illustration of the power of the variational method. The trochoidal wave described here is probably the approximation with the highest ratio {\em accuracy\//\/complexity\/} ever derived for a traveling wave in deep water. This approximation has first been derived by \textsc{Geniet} \cite{Geniet2003} via a different approach.

\section{Arbitrary depth}
\label{secfd}

A general ansatz, for waves in finite constant depth and satisfying identically the bottom impermeability, is suggested by the linear theory of water waves:
\begin{align}\label{ansfin}
  \phi \approx \frac{\cosh\kappa\/Y}{\cosh\kappa\/h} \phis(\vx,t),
  \quad \vu \approx \frac{\cosh\kappa\/Y}{\cosh\kappa\/h}\vus(\vx,t), \quad v \approx
  \frac{\sinh\kappa\/Y}{\sinh\kappa\/h} \vs(\vx,t), \nonumber \\
  \vmu \approx \frac{\cosh\kappa\/Y}{\cosh\kappa\/h}\vmus(\vx,t), \quad \nu \approx \frac{\sinh\kappa\/Y}{\sinh\kappa\/h} \nus(\vx,t),
\end{align}
where $Y = y + d$ and $h = \eta + d$. The parameter $\kappa$ is a characteristic wave number to be made precise {\em a posteriori\/}. This ansatz is uniformly valid for all depths because it yields the shallow water one (\ref{anssha}) as $\kappa\rightarrow0$, and the deep water one (\ref{ansinf}) as $d\rightarrow\infty$. Obviously, the ansatz (\ref{ansfin}) is valid for wave fields with wavenumber spectra that are narrow-banded around $\kappa$. Substituting the ansatz (\ref{ansfin}) into the relaxed variational principle (\ref{defL3}), we obtain
\begin{align}\label{lagfin}
  \mathscr{L}\ =&\ [\,\eta_t\,+\,\vmus\scal\nab\eta\,]\,\phis\ -\ \half\,g\,\eta^2\ +\ [\,\nus\,\vs\,-\,\half\,\vs^2\,]\,\frac{\sinh(2\kappa h)-2\kappa h}{2\kappa\cosh(2\kappa h) - 2\kappa} \nonumber \\
  &+\ [\/\vmus\scal\vus-{\half}\/\vus^2+\phis\nab\scal\vmus-\kappa\tanh(\kappa h)\phis\,\vmus\scal\nab\eta\/]\,\frac{\sinh(2\kappa h)+2\kappa h}{2\kappa\cosh(2\kappa h)+2\kappa} \nonumber \\
  & +\ \half\,\phis\,\nus\left[\,\frac{2\kappa h}{\sinh(2\kappa h)}\,-\,1\,\right].
\end{align}

Applying various constraints, one obtains generalized equations including the ones derived in Sections~\ref{secsw} and \ref{secdw} as limiting cases. In particular, we can derive arbitrary depth generalizations of the Saint-Venant and `irrotational' Green--Naghdi equations. Thus, the relaxed variational principle leads simple derivations of model equations for finite amplitude waves in arbitrary depth, which is not the case with the classical perturbations techniques. These developments are left to future investigations. We note also some similarity between the ansatz (\ref{ansfin}) and the coupled-mode theory for small-amplitude water waves \cite{Athanassoulis1999}.

\section{Generalizations}\label{secgenans}

The ansatz (\ref{ansfin}) can be generalized in many relevant ways, depending on the problem under consideration. Natural generalizations can be based on the velocity vertical variations given by, \eg, higher-order deep and shallow water theories, or obtained fitting some experimental data. In this section we propose a possible generalization based on an ansatz of the form
\begin{equation}\label{ansfingen}
  \phi \approx \left[\frac{\cosh\kappa\/Y}{\cosh\kappa\/h}\right]^\lambda\phis(\vx,t), \quad 
  \vu \approx \left[\frac{\cosh\kappa\/Y}{\cosh\kappa\/h}\right]^\lambda\vus(\vx,t), \quad 
  v \approx \left[\frac{\sinh\kappa\/Y}{\sinh\kappa\/h}\right]^\lambda\vs(\vx,t),
\end{equation}
where $\lambda$ is a parameter at our disposal. If $\lambda = 1$, the ansatz (\ref{ansfin}) is recovered, but the case $\lambda \neq 1$ does not correspond to the vertical profile predicted by any theory based on perturbation expansions. Still, this type of ansatz is of some interest, as we shall see below.

Note first that in the deep water limit $d\rightarrow\infty$, the ansatz (\ref{ansfingen}) becomes
\begin{equation*}
  \phi\ \approx\, \ue^{\lambda\kappa\/(y-\eta)} \phis(\vx,t), \quad
  \vu\ \approx\ \ue^{\lambda\kappa\/(y-\eta)}\,\vus(\vx,t), \quad
  v\ \approx\ \ue^{\lambda\kappa\/(y-\eta)}\, \vs(\vx,t),
\end{equation*}
and thus, via the change of parameter $\lambda\kappa\mapsto\kappa$, the ansatz (\ref{ansinf}) is recovered. This means that (\ref{ansfingen}) is not more general than (\ref{ansfin}) in deep water. On the contrary, these two ansatz are very different in finite depth. We illustrate this claim in the simple case of shallow water ($\kappa d\rightarrow 0$) when (\ref{ansfingen}) becomes
\begin{align*}
  \phi\ \approx\ \phim(\vx,t), \qquad
  \vu\ \approx\ \vum(\vx,t), \qquad
  v\ \approx\,\left[\,\frac{y+d}{\eta+d}\,\right]^\lambda\, \vs(\vx,t),
\end{align*}
where we have replaced $\phis$ and $\vus$ by $\phim$ and $\vum$, respectively, since they are equal in this limiting case.

\subsection{Modified Serre's equations}

For the sake of simplicity, we consider here only one horizontal dimension, say $x_1$, and we set $x_1 = x$ and $u_1 = u$, for brevity. We also consider the special case $\mu = u$ and $\nu = v$ together with the constraint $\vs = \eta_t + \us\/\eta_x$ (free surface's impermeability). Thus, the Lagrangian density (\ref{defL3}) yields
\begin{equation}\label{lagdensermod}
  \mathscr{L}\, =\, \left(\/\eta_t\/+\/[\/(\eta+d)\/\um\/]_x\/\right)\phis\ -\ \half\/g\/\eta^2\, +\, \half\/(\eta+d)\/\um^2\, +\, \half\/\beta^2\/(\eta+d)\left[\/\eta_t\/+\/\um\/\eta_x\/\right]^2,
\end{equation}
where $\beta^2 = (2\lambda+1)^{-1}$. After some algebra, the Euler--Lagrange equations leads to the following equations
\begin{eqnarray}\label{eqsermod}
  h_t\ +\ [\,h\,\um\,]_x\ &=\ 0, \\
  \um_t\ +\ \um\,\um_x\ +\ g\,h_x\ +\ \beta^2\,h^{-1}\,[\,h^2\,\tilde{\gamma}\,]_x\ &=\ 0, \label{eqsermod2}
\end{eqnarray}
where $\tilde{\gamma}$ is defined in (\ref{eqgamma_tilde}). If $\beta = 1/\sqrt{3}$ the classical Serre equations are recovered.

Equations (\ref{eqsermod})--(\ref{eqsermod2}) admit a ($2\pi/k$)-periodic cnoidal traveling wave solution:
\begin{align}\label{solsermod}
  \um\ =\ \frac{c\,\eta} {d+\eta},\qquad 
  \eta\ =\ a\,\frac{\operatorname{dn}^2\!\left(\/\half\/\kappa\/\xi\/|\/m\/\right)-E/K} {1-E/K}\ =\ a\, -\, H\,\operatorname{sn}^2\!\left(\/\half\/\kappa\/\xi\/|\/m\/\right),
\end{align}
with $\xi = c - ct$, $\text{dn}$ and $\text{sn}$ being elliptic functions of Jacobi of parameter $m$ ($0\leqslant m\leqslant1$), and where $K = \text{K}\/(m)$ and $E = \text{E}\/(m)$ are the complete ellipic integrals of the first and second kinds, respectively (\textsc{Abramowitz} \& \textsc{Stegun} 1965, \#17.3, \cite{Abramowitz1965}). The parameter $\kappa$ is a sort of wave\-number, $a$ is the wave amplitude (mean level to crest elevation), $H$ is the total wave height (trough to crest elevation) and $c$ is the wave phase velocity observed in the frame of reference without mean flow. The wave parameters are related via the relations
\begin{eqnarray}\label{relparasolsermod}
  k\ =\ \frac{\pi\,\kappa}{2\,K}, \qquad  H\ =\ \frac{m\,a\,K}{K-E}, \qquad
  (\kappa\/d)^2\ =\ \frac{g\,H}{m\,\beta^2\,c^2},\\
  \gdef\thesubequation{\theequation\textit{d}} 
  m\ =\ \frac{g\,H\,(d+a)\,(d+a-H)}{g\,(d+a)^2\,(d+a-H)\,-\,d^2\,c^2}.
  \label{relparasolsermod2}
\end{eqnarray}
In the limiting case $m\rightarrow1$, we have $K\rightarrow\infty$, $E/K\rightarrow0$, $k\rightarrow0$, $H\rightarrow a$ and hence, the classical solitary wave solution is recovered
\begin{eqnarray*}
  \eta\ =\ a\,\text{sech}^2\half\/\kappa\/(x-ct), \qquad
  c^2\ =\ g\,(d+a), \qquad 
  \frac{a}{d}\ =\ \frac{(\beta\/\kappa\/d)^2}{1\,-\,(\beta\/\kappa\/d)^2}. \qquad\qquad
\end{eqnarray*}

At this stage, $\beta$ is still a free parameter. An suitable expression for this parameter may be obtained substituting the solution (\ref{solsermod}) into the Lagrangian density (\ref{lagdensermod}), integrating $\mathscr{L}$ over one wavelength, then solving $\ud\mathcal{L}/\ud\beta = 0$ keeping $k$ and $c$ constant (as well as $g$ and $d$), the other parameters varying according to relations (\ref{relparasolsermod}-\ref{relparasolsermod2}). Thus, after some cumbersome algebra, we found $\beta = 0$ for this parameter, which is not very interesting for practical applications. A possible alternative here is to choose $\beta$ such that the exact relation $c^2 = g\tan(\kappa d)/\kappa$ is satisfied identically or up to some asymptotic order.

%%% ------------------------------------------------------------------------ %%%

\section{Discussion}
\label{concl}

In this study, we have illustrated the advantage of using a variational principle with as many variables as possible. We call it the {\em relaxed\/} variational principle, since the Lagrangian density (\ref{defL3}) involves more degrees of freedom (\ie, the variables $\eta$, $\phi$, $\vu$, $v$, $\vmu$ and $\nu$) compared to the two degrees of freedom ($\eta$ and $\phi$) in the classical case. In particular, these extra variables can be used to impose various constraints such as incompressibility, irrotationality, impermeability, etc. The practical use of the relaxed formulation was illustrated on numerous examples in shallow, deep and intermediate waters. Thus, we obtained several approximations, some well-known, some new to our knowledge. 

In the shallow water regime, we have first obtained the classical nonlinear shallow water (or Saint-Venant) equations (\ref{eqswsv})--(\ref{eqswsv2}). Then, with the same ansatz (\ref{anssha}) but imposing the constraint of the free surface impermeability, we have derived the irrotational Green--Naghdi equations (\ref{eqswser})--(\ref{eqswser2}). Applying the incompressibility constraint and choosing differently the pseudo-velocity field, we have obtained two kinds of {\em generalized Kaup--Boussinesq} equations. Several ways of further generalizations were also outlined.

In deep water, two models were considered. Namely, we derived deep water counterparts of the celebrated Saint-Venant and Serre equations. The former has a canonical Hamiltonian formulation and degenerates to the Klein--Gordon equation in the linear approximation; we thus called the new system (\ref{gkg})--(\ref{surisoL21}) {\em generalized Klein--Gordon\/} equations. The latter could be solved analytically for a two-dimensional traveling wave. This solution is a striking illustration of the power of the variational formulation compared to asymptotic expansion methods, especially for large amplitudes when the expansion parameter is no longer small. In addition, both equations were shown to possess excellent asymptotic properties with respect to Stokes-like expansions.

The case of arbitrary depth has also been briefly considered. In particular, it has been shown how easily one can introduce an ansatz valid for all depths. Indeed, the vertical variation of the velocity field suggested by the linear theory provides at once such a general ansatz which degenerates to previous cases when the water is shallow ($\kappa d\to 0$) or deep ($d\to\infty$). This simplicity and flexibility of the variational principle is quite remarkable compared to perturbation methods.

Finally, we considered a generalized ansatz and we illustrated its consequence in the limiting case of shallow water. In this way, we derived the {\em modified Serre equations} and subsequently obtained exact cnoidal and solitary wave solutions. The main purpose of this example was to illustrate the fact that one can introduce an ansatz which is not inspired by any asymptotic expansion and nevertheless lead to reasonable approximations.

In the present paper, some further possibilities for generalizations are also mentioned. However, we have to emphasize that not all ansatz and constraints will necessarily lead to physically relevant and tractable approximations; the same is true for models derived from asymptotic expansions, however. Nonetheless, the relaxed variational formulation is sufficiently versatile to allow easy derivations of physically sound models. We have illustrated this claim, in particular, by showing how it is simple to obtain approximate equations valid for all depths.

Sometimes, the choice of the constraints may seem to be rather {\em ad hoc}, but that should not be surprising. Indeed, the water wave theory already knows several {\em ad hoc\/} `tricks' intended to improve the approximation quality. For instance, it was proposed in \cite{Madsen03} to replace the polynomial shallow water expansion (\ref{eq:swe}) by a $(m,n)$-Pad\'e approximation, the orders $m$ and $n$ being chosen to improve the linear dispersion relation of progressive waves. Another example is the use of the velocity potential defined at some depth $y_0$ \cite{Nwogu1993} and, as before, the free parameter $y_0$ is chosen to improve linear dispersion characteristics. The approach proposed here is not more {\em ad hoc\/} that any example mentioned above. Moreover, the variational principle allows for greater flexibility in the choice of ansatz. Thus, the approximations derived via the relaxed variational procedure must be studied {\em a posteriori} in order to verify their mathematical consistence and their physical relevance. This is also the case for approximations derived via perturbation techniques, and many such approximations commonly used have not yet been justified on a rigorous mathematical basis. 

In the several examples presented here, the ansatz involve free parameters that we have chosen constant for simplicity. One may also consider these parameters as functions and find their values from the stationary point of the Lagragian. Doing so will lead to more complicated equations, but this is not a major issue if these equations are intended to be treated numerically.

In order to derive approximate models, variational formulations are attractive alternative to asymptotic expansions. However, both approaches can be also combined. Indeed, once the variational principle has been applied to an ansatz, asymptotic expansions can be further applied to obtain simpler models. For instance, one could consider `unidirectionalized' approximations \cite{Olver1984, Olver1988a} to derive variants of \textsc{Korteweg} and \textsc{de Vries} \cite{KdV}, \textsc{Dysthe} \cite{Dysthe1979}, Camassa and Holm \cite{Camassa1993}, \textsc{Degasperis} and \textsc{Procesi} \cite{Degasperis1999}, \textsc{Kraenkel} \etal \cite{Kraenkel2005}, and other equations. This possibility will be investigated in future works.
 
For the sake of simplicity, we have considered only gravity waves propagating at the surface of a single layer of a homogeneous fluid with a horizontal bottom. It is trivial to introduce a relaxed variational formulation including, \eg, surface tension, stratifications in several homogeneous layers and obstacles. Such general variational formulations, together with relevant ansatz and well chosen constraints, will easily lead to interesting models. For perfect fluids, variational formulations can also be obtained for rotational motions \cite{Eckart1960, Luke1967, Lewis1986, Salmon1988, Morrison1998, Constantin2006}. A relaxed version of such variational principles will facilitate the derivation of approximate models.

The numerical models for simulating water waves are undergoing constant improvements. The state of the art can be found in recent reviews (\eg, \cite{Fenton1999, Dias2006a, Ma2010}). Certainly, the variational principle is the tool of choice to derive efficient approximations. By efficient, we mean models that capture most of the relevant physics and which, in the same time, can be easily and rapidly solved numerically. One interesting direction for future researches is the development of numerical schemes preserving hamiltonian structure at the discrete level. Another interesting application is the derivation of new improved models with uneven bathymetry for coastal hydrodynamics and tsunami wave modeling (\textsc{Synolakis} \& \textsc{Bernard} 2006, \cite{Syno2006}).

The use of a variational principle for modeling surface waves is by no mean new. However, its power has not yet been fully exploited. The present paper is a further contribution in this direction. Obviously, advantages of relaxed variational principles may as well be exploited in physical contexts other than water waves, such as plasma physics, nonlinear optics, etc.

%%% ------------------------------------------------------------------------ %%%

\subsection*{Acknowledgement}
\addcontentsline{toc}{subsection}{Acknowledgement}

The second author would like to acknowledge the support from ANR MathOc\'ean (Project n$^\circ$ ANR-08-BLAN-0301-01) and from the program ``Risques gravitaires, s\'eismes'' of Cluster Environement and the research network VOR.

%%% ------------------------------------------------------------------------ %%%

\appendix
\section{Exact Stokes wave}
\label{appsto}

In deep water, a seventh-order Stokes expansion (for the exact equations) is
\begin{eqnarray*}
  \kappa\,\eta&=&\alpha\cos\theta\,+\,{\textstyle{1\over2}}\alpha^2\!\left(1\!+\!{\textstyle{17\over12}}\alpha^2\!+\!{\textstyle{233\over64}}\alpha^4\right)\cos2\theta \qquad \nonumber \\
  && +\,{\textstyle{3\over8}}\alpha^3\!\left(1\!+\!{\textstyle{51\over16}}\alpha^2\!+\!{\textstyle{3463\over320}}\alpha^4\right)\cos3\theta\,+\,{\textstyle{1\over3}}\alpha^4\!\left(1\!+\!{\textstyle{307\over60}}\alpha^2\right)\cos4\theta  \\
  &&+\,{\textstyle{125\over384}}\alpha^5\!\left(1\!+\!{\textstyle{10697\over1500}}\alpha^2\right)\cos5\theta\,+\,{\textstyle{27\over80}}\alpha^6\cos6\theta\,+\,{\textstyle{16807\over46080}}\alpha^7\cos7\theta\,+\,\mathrm{O}(\alpha^8), \nonumber \\
  {g^{-{1\over2}}}\/\kappa^{3\over2}\,\phis&=&\alpha\left(1\!-\!{\textstyle{1\over4}}\alpha^2\!-\!{\textstyle{43\over96}}\alpha^4\!-\!{\textstyle{2261\over1536}}\alpha^6\right)\sin\theta\,+\,{\textstyle{1\over2}}\alpha^2\!\left(1\!+\!{\textstyle{7\over12}}\alpha^2\!+\!{\textstyle{81\over64}}\alpha^4\right)\sin2\theta \nonumber \\
  && +\,{\textstyle{3\over8}}\alpha^3\!\left(1\!+\!{\textstyle{281\over144}}\alpha^2\!+\!{\textstyle{5813\over1080}}\alpha^4\right)\sin3\theta\,+\,{\textstyle{1\over3}}\alpha^4\!\left(1\!+\!{\textstyle{431\over120}}\alpha^2\right)\sin4\theta  \\
  &&+\,{\textstyle{125\over384}}\alpha^5\!\left(1\!+\!{\textstyle{3369\over625}}\alpha^2\right)\sin5\theta\,+\,{\textstyle{27\over80}}\alpha^6\sin6\theta\,+\,{\textstyle{16807\over46080}}\alpha^7\sin7\theta\,+\,\mathrm{O}(\alpha^8), \nonumber \\
  {g^{-{1\over2}}}\/\kappa^{1\over2}\,c &=& 1+{\textstyle{1\over2}}\alpha^2+{\textstyle{1\over2}}\alpha^4+{\textstyle{707\over384}}\alpha^6+\mathrm{O}(\alpha^8),
\end{eqnarray*}
where $\theta = \kappa(x-ct)$. Note that, to the leading order, the $n$-th Fourier coefficient is $2^{1-n}n^{n-2}\alpha^n/(n\!-\!1)!$ (this is also true for all $n>7$). In the bulk of the fluid, the velocity potential is
\begin{eqnarray*}
  {g^{-{1\over2}}}\/\kappa^{3\over2}\,\phi&=&\alpha\left(1\!-\!{\textstyle{1\over8}}\alpha^2\! - \!{\textstyle{7\over12}}\alpha^4\!-\!{\textstyle{14761\over9216}}\alpha^6\right)\ue^{\kappa\/y}\sin\theta\, + \,{\textstyle{1\over2}}\alpha^4\!\left(1\!+\!{\textstyle{11\over6}}\alpha^2\right) \ue^{2\kappa\/y}\sin2\theta \nonumber \\
  && +\,{\textstyle{1\over12}}\alpha^5\!\left(1\!+\!{\textstyle{191\over24}}\alpha^2\right)\ue^{3\kappa\/y}\sin3\theta\,+\,{\textstyle{1\over72}}\alpha^6\,\ue^{4\kappa\/y}\sin4\theta  \\
  &&+\,{\textstyle{1\over480}}\alpha^7\,\ue^{5\kappa\/y}\sin5\theta\,+\,\mathrm{O}(\alpha^8), \nonumber
\end{eqnarray*}
meaning that harmonics appear at the fourth-order only, thus justifying the ansatz (\ref{ansinf}). Note that, to the leading order, the $n$-th Fourier coefficient is $\alpha^{n+2}/n!(n\!-\!1)$ for all $n > 1$.

\section{Cubic Zakharov's equations} \label{appcze}

Satisfying exactly the Laplace equation and the bottom impermeability, the gravity waves variational formulation \cite{Luke1967} yields the Hamiltonian \cite{Zakharov1968}:
\begin{eqnarray} \label{Hexa}
  \mathcal{H} &=& \half\int\left\{\,g\,\eta^2\,+\,\phis\,V\,\right\}\,\ud^2\/\vx, \qquad V=\left[\,\phi_y-\nab\eta\scal\nab\phi\,\right]_{y=\eta}.
\end{eqnarray}
Introducing a Dirichlet--Neumann operator $G$, such that $V=G(\eta)\phis$ (Craig \& Sulem 1993, \cite{Craig1993}), expanding  $G$ around $\eta=0$ and neglecting the terms beyond the quartic nonlinearities, the Hamiltonian (\ref{Hexa}) becomes
\begin{eqnarray} \label{Hzak}
  \mathcal{H} &=& \half\int\left\{ g\/\eta^2+\phis\left[\,\der\phis-\der(\eta\/\der\phis)- \nab\scal(\eta\/\nab\phis)\right.\right. \nonumber \\
  &&\hspace{10mm}+\left.\left.\half\/\der(\eta^2\/\nab^2\phis)+\der(\eta\,\der(\eta\,\der\phis)) + \half\/\nab^2(\eta^2\,\der\phis)\,\right]\right\}\,\ud^2\/\vx,
\end{eqnarray}
with the pseudo-differential operator $\der = (-\nab^2)^{1\over2} \tanh[(-\nab^2)^{1\over2}d\,]$. (For one horizontal dimension in infinite depth $\der f=-\mathfrak{H}(f_x)$, $\mathfrak{H}$ the Hilbert transform.) Thus, the cubic Zakharov's equations (CZE) are
\begin{align}\label{eqcze}
\quad\eta_t-\der\phis\, &=\, -\nab\scal(\eta\nab\phis) - \der(\eta\der\phis) + %\nonumber \\
 \half\nab^2(\eta^2\der\phis) + \der(\eta\der(\eta\der\phis)) + 
\half\der(\eta^2\nab^2\phis), \\
\phis_t + g\eta\, &=\, \half(\der\phis)^2 - \half(\nab\phis)^2 - (\eta\der\phis)\nab^2\phis - 
(\der\phis)\der(\eta\der\phis). \label{eqcze2}
\end{align}

For progressive ($2\pi/\kappa$)-periodic solutions in infinite depth, a seventh-order Stokes expansion is
\begin{eqnarray*}
\kappa\,\eta&=&\alpha\cos\theta
 \,+\,{\textstyle{1\over2}}\alpha^2\!\left(1\!+\!{\textstyle{{\bf3}\over{\bf2}}}\alpha^2\!
 +\!{\textstyle{{\bf445}\over{\bf96}}}\alpha^4\right)\cos2\theta \nonumber \\
&& +\,{\textstyle{3\over8}}\alpha^3\!\left(1\!+\!
{\textstyle{{\bf41}\over{\bf12}}}\alpha^2\!+\!{\textstyle{{\bf5213}\over{\bf384}}}\alpha^4\right)\cos3\theta
\,+\,{\textstyle{\bf7\over24}}\alpha^4\!\left(1\!+\!
{\textstyle{{\bf263}\over{\bf42}}}\alpha^2\right)\cos4\theta \\
&&+\,{\textstyle{{\bf67}\over384}}\alpha^5\!\left(1\!+\!
{\textstyle{{\bf2569}\over{\bf201}}}\alpha^2\right)\cos5\theta \fatminus {\textstyle{\bf
9\over320}}\alpha^6\cos6\theta
\fatminus {\textstyle{{\bf16751}\over46080}}\alpha^7\cos7\theta\,+\,\mathrm{O}(\alpha^8), \nonumber \\
{g^{-{1\over2}}}\/\kappa^{3\over2}\,\phis&=&\alpha\left(1\!-\!{\textstyle{1\over4}}\alpha^2\!
 -\!{\textstyle{{\bf31}\over{\bf64}}}\alpha^4\!-\!{\textstyle{{\bf465}\over{\bf256}}}
 \alpha^6\right)\sin\theta
 \,+\,{\textstyle{1\over2}}\alpha^2\!\left(1\!+\!{\textstyle{{\bf3}\over{\bf4}}}\alpha^2\!
 +\!{\textstyle{{\bf123}\over{\bf64}}}\alpha^4\right)\sin2\theta \nonumber \\
&& +\,{\textstyle{3\over8}}\alpha^3\!\left(1\!+\!
{\textstyle{{\bf89}\over{\bf36}}}\alpha^2\!+\!{\textstyle{{\bf27271}\over{\bf3456}}}
\alpha^4\right)\sin3\theta
\,+\,{\textstyle{\bf7\over24}}\alpha^4\!\left(1\!+\!
{\textstyle{{\bf1795}\over{\bf336}}}\alpha^2\right)\sin4\theta  \\
&&+\,{\textstyle{{\bf67}\over384}}\alpha^5\!\left(1\!+\!
{\textstyle{{\bf24769}\over{\bf2010}}}\alpha^2\right)\sin5\theta
\fatminus{\textstyle{\bf9\over320}}\alpha^6\sin6\theta
\fatminus{\textstyle{{\bf16751}\over46080}}\alpha^7\sin7\theta\,+\,\mathrm{O}(\alpha^8), \nonumber  \\
{g^{-{1\over2}}}\/\kappa^{1\over2}\,c &=& 1+{\textstyle{1\over2}}\alpha^2
+{\textstyle{\bf41\over64}}\alpha^4+{\textstyle{{\bf913}\over384}}\alpha^6+\mathrm{O}(\alpha^8),
\end{eqnarray*}
where the incorrect (compared to the exact expansion) coefficients and signs are displayed in bold face. Thus, the CZE match the exact Stokes wave up to the third-order only. Truncating the Hamiltonian at the order $N + 1$ in nonlinearities, the corresponding Stokes double series is correct up to the order $N$ in the expansion parameter. None of these approximations have the exact asymptotic behavior (\ref{fourcoef}) for their Fourier coefficients.

%%% ------------------------------------------------------------------------ %%%

%%% Bibliography %%%
\addcontentsline{toc}{section}{References}
\bibliographystyle{abbrv}
\bibliography{biblio}

\end{document}